\documentclass[sigconf]{acmart}

\usepackage{amsmath}
\usepackage{courier}
\usepackage{bm}
\usepackage{algpseudocode}
\usepackage[noend,ruled,linesnumbered]{algorithm2e}
\usepackage{color}
\usepackage{enumitem}
\usepackage{url}
\usepackage{graphicx}
\usepackage{array}
\usepackage{multirow}
\usepackage{pifont}
\usepackage{xcolor,colortbl}
\usepackage{setspace}
\usepackage{subfig}
\usepackage{graphbox}
\usepackage{makecell}
\usepackage{nicematrix}
\usepackage{dblfloatfix}
\usepackage{balance}

\newcommand{\dataset}{Maple}
\newcommand{\bmx}{\bm x}
\newcommand{\bmy}{\bm y}
\newcommand{\bmv}{\bm v}
\newcommand{\bmh}{\bm h}
\newcommand{\bmH}{\bm H}
\newcommand{\bmu}{\bm u}

\definecolor{myblue}{rgb}{0.2, 0.2, 0.9}

\newcommand\blfootnote[1]{%
  \begingroup
  \renewcommand\thefootnote{}\footnote{#1}%
  \addtocounter{footnote}{-1}%
  \endgroup
}

\copyrightyear{2023}
\acmYear{2023}
\setcopyright{acmlicensed}\acmConference[WWW '23]{Proceedings of the ACM Web Conference 2023}{May 1--5, 2023}{Austin, TX, USA}
\acmBooktitle{Proceedings of the ACM Web Conference 2023 (WWW '23), May 1--5, 2023, Austin, TX, USA}
\acmPrice{15.00}
\acmDOI{10.1145/3543507.3583354}
\acmISBN{978-1-4503-9416-1/23/04}

\begin{document}
\title{The Effect of Metadata on Scientific Literature Tagging: \\
A Cross-Field Cross-Model Study}

\author{Yu Zhang, Bowen Jin, Qi Zhu, Yu Meng, Jiawei Han}
\affiliation{
\institution{University of Illinois at Urbana-Champaign}
\country{\{yuz9, bowenj4, qiz3, yumeng5, hanj\}@illinois.edu}
}

\renewcommand{\shortauthors}{Zhang et al.}

\begin{spacing}{0.95}

\begin{abstract}
Due to the exponential growth of scientific publications on the Web, there is a pressing need to tag each paper with fine-grained topics so that researchers can track their interested fields of study rather than drowning in the whole literature. Scientific literature tagging is beyond a pure multi-label text classification task because papers on the Web are prevalently accompanied by metadata information such as venues, authors, and references, which may serve as additional signals to infer relevant tags. Although there have been studies making use of metadata in academic paper classification, their focus is often restricted to one or two scientific fields (e.g., computer science and biomedicine) and to one specific model. In this work, we systematically study the effect of metadata on scientific literature tagging across 19 fields. We select three representative multi-label classifiers (i.e., a bag-of-words model, a sequence-based model, and a pre-trained language model) and explore their performance change in scientific literature tagging when metadata are fed to the classifiers as additional features. We observe some ubiquitous patterns of metadata's effects across all fields (e.g., venues are consistently beneficial to paper tagging in almost all cases), as well as some unique patterns in fields other than computer science and biomedicine, which are not explored in previous studies.
\blfootnote{$^\dagger$Code and Datasets are available at {\color{myblue} \url{https://github.com/yuzhimanhua/MAPLE}} and {\color{myblue} \url{https://doi.org/10.5281/zenodo.7611544}}, respectively.}
\end{abstract}

\ccsdesc[500]{Information systems~Digital libraries and archives}
\ccsdesc[500]{Information systems~Data mining}
\ccsdesc[500]{Information systems~World Wide Web}

\keywords{scientific literature tagging; metadata; text classification}

\maketitle

\section{Introduction}
\begin{figure}[t]
\centering
\includegraphics[width=0.95\linewidth]{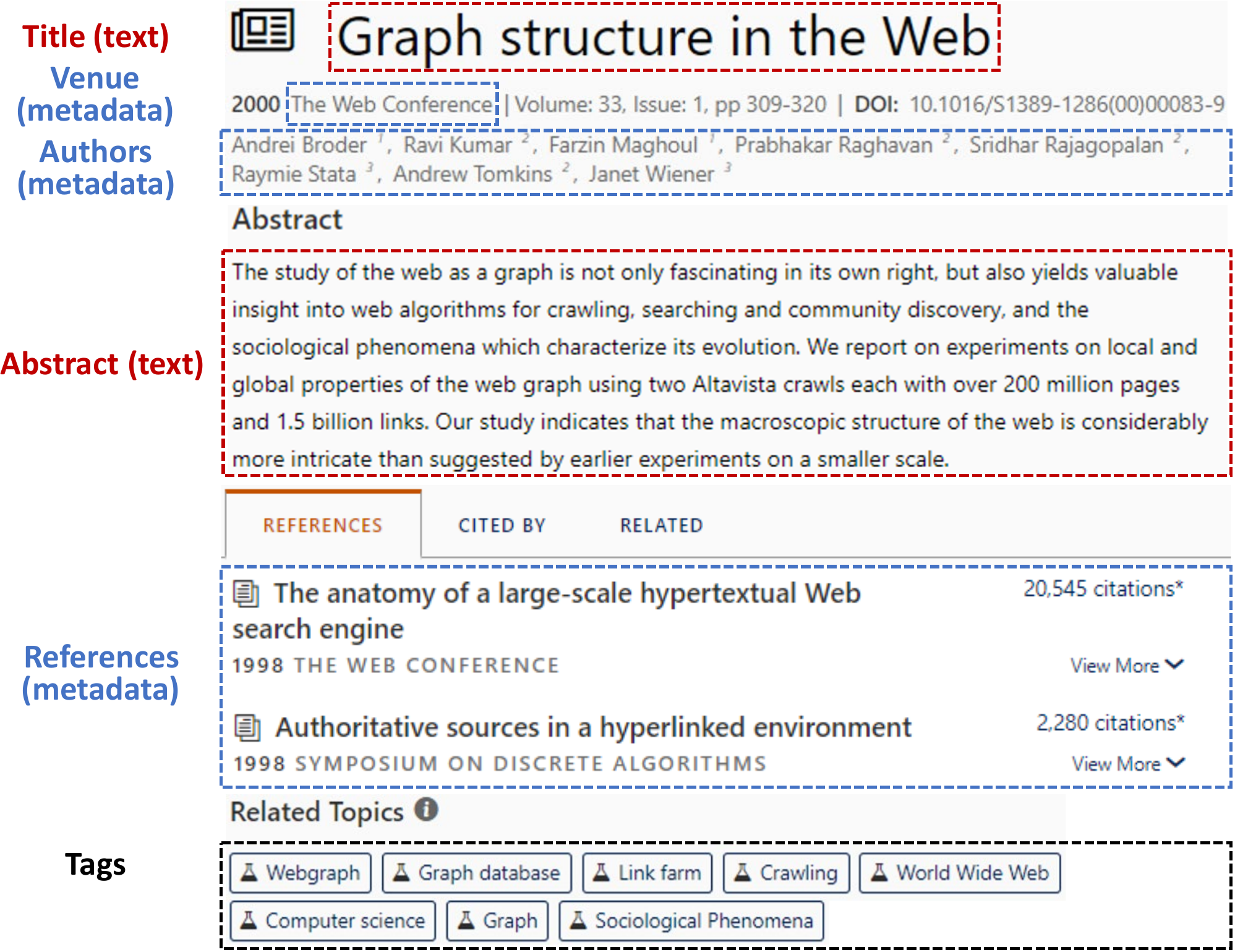}
\vspace{-0.5em}
\caption{A scientific paper with metadata from Microsoft Academic \cite{sinha2015overview}. The goal of scientific literature tagging is to predict its related topics.}
\vspace{-1em}
\label{fig:intro}
\end{figure}

A variety of academic service platforms, such as Google Scholar, AMiner \cite{tang2008arnetminer}, Microsoft Academic \cite{sinha2015overview}, Semantic Scholar \cite{ammar2018construction}, and PubMed \cite{lu2011pubmed}, are available on the Web with great attention received. One major goal of these platforms is to help researchers query and track academic information and resources. Meanwhile, the volume of scientific publications is growing exponentially, doubling every 12 years \cite{dong2017century} and reaching 240,000,000 by 2019 \cite{wang2020microsoft}. In such an information explosion era, it becomes more important than ever to accurately tag each scientific paper with its relevant topics so that researchers can track their interested fields of study instead of getting overwhelmed by the whole literature. Figure \ref{fig:intro} shows an example of the scientific literature tagging task, which aims to predict the tags such as ``\textsf{World Wide Web}'', ``\textsf{Webgraph}'', and ``\textsf{Link Farm}'' given the paper ``\textit{Graph structure in the Web}''.

Previous studies \cite{zhang2021match,ye2021beyond} have pointed out that scientific literature tagging is beyond a multi-label text classification task because academic papers are accompanied by metadata, which make them more complex than plain text sequences. Such metadata information, including venues, authors, and references, can be strong indicators of each paper's related topics. For example, in Figure \ref{fig:intro}, the venue ``\textit{WWW}'' implies the paper's relevance to ``\textsf{Computer Science}'' and ``\textsf{World Wide Web}'', while the authors and references may further indicate fine-grained tags such as ``\textsf{Webgraph}'' and ``\textsf{Link Farm}''.

Although existing studies on scientific literature tagging \cite{xun2019meshprobenet,zhang2021match,ye2021beyond,zhang2022metadata,zhang2021hierarchical} have proposed to incorporate metadata features into the tagger, they still have two major limitations. First, their focus is restricted to one or two scientific fields only (e.g., computer science and biomedicine). Empirically examining the effect of metadata in other fields (e.g., art, economics, mathematics, physics, etc.) has remained elusive, mainly owing to the lack of benchmark datasets for fine-grained paper tagging in these fields. Second, the proposed metadata-aware taggers mainly train an RNN \cite{chung2014empirical} or Transformer \cite{vaswani2017attention} architecture from scratch. It is still unclear whether the proposed usage of metadata can be generalized to other approaches, such as bag-of-words classifiers (which are still broadly studied in large-scale multi-label text classification when efficiency is concerned \cite{liu2021emerging}) and pre-trained language models.

\vspace{1mm}

\noindent \textbf{Contributions.} To address the aforementioned two limitations of previous studies, in this work, we conduct a systematic \emph{cross-field} \emph{cross-model} study on the effect of metadata on scientific literature tagging. First, we construct a large-scale scientific literature tagging benchmark, \textsc{\dataset} (\underline{M}etadata-\underline{A}ware \underline{P}aper col\underline{LE}ction), 
from the Microsoft Academic Graph \cite{sinha2015overview}. \textsc{\dataset} covers 19 scientific fields and consists of more than 11.9 million papers. The number of candidate tags in each field ranges between $\sim$700 and $\sim$64,000. Then, we consider three major types of multi-label classification approaches: bag-of-words classifiers \cite{prabhu2018parabel,babbar2017dismec,yen2017ppdsparse,prabhu2018extreme,jain2016extreme,prabhu2014fastxml,tagami2017annexml,yen2016pd,peng2016deepmesh,yu2022pecos,khandagale2020bonsai}, sequence-based classifiers \cite{liu2017deep,xun2019meshprobenet,you2019attentionxml,zhang2021match,ye2021beyond,xun2020correlation}, and pre-trained language models \cite{chang2020taming,you2021bertmesh,zhang2022metadata,jiang2021lightxml,ye2020pretrained,zhang2021fast}. We select one representative model from each of the three categories, namely Parabel \cite{prabhu2018parabel}, Transformer \cite{xun2020correlation}, and OAG-BERT \cite{liu2022oag}, that can be modified in a straightforward way to jointly take text and metadata as input for classification. Based on \textsc{\dataset}, we explore the effect of venues, authors, and references on paper tagging in the 19 fields when using the three selected classifiers. We have the following major observations:

\begin{itemize}[leftmargin=*]
\item The effect of metadata varies significantly across different fields and classifiers. In general, venues are consistently beneficial in almost all cases, while the benefit of authors and references is highly dependent on the field and the classifier. For example, author information is evidently beneficial to paper tagging in the Philosophy field while harmful in Geology; references are helpful in Mathematics when we use Parabel but become disadvantageous in the same field when Transformer is adopted.
\item The effect of metadata tends to be similar in two fields that belong to the same high-level scientific area. For example, Biology and Medicine are both life sciences (according to \cite{rosvall2011multilevel,yin2022public}), and the effects of venues, authors, and references are largely aligned in the two fields. This finding implies that the experience of using metadata in one field can be extrapolated to a similar field.
\item We also study the effect of metadata when predicting tags at different granularity levels. In a number of fields, venues improve the performance for not only coarse-grained tags but also very fine-grained ones, which may be unusual in the Computer Science field. The major reason is that some venues in these fields can indicate very fine tags (e.g., ``\textit{Journal of Roman Archaeology}'' in History, ``\textit{Mediaeval Studies}'' in Philosophy).
\end{itemize}

To summarize, this work makes the following contributions: (1) We construct a large-scale benchmark, \textsc{\dataset}, for scientific literature tagging across 19 fields, whose field coverage is much broader than the datasets used in previous paper tagging studies \cite{xun2019meshprobenet,zhang2021match,ye2021beyond}. (2) We comprehensively evaluate the performance of different types of multi-label classifiers in scientific literature tagging after incorporating metadata features. (3) Our empirical findings demonstrate some ubiquitous patterns of metadata's effects across all fields, as well as some unique patterns in fields other than computer science and biomedicine, which are not explored in previous studies. 
Our systematic studies are meant for providing insights to practitioners to build scientific literature taggers that can benefit all fields.

\section{Preliminaries}
\noindent \textbf{Text and Metadata.} We represent the text information of a paper $p$ as a single text sequence $\mathcal{T}_p=w_1w_2...w_{|\mathcal{T}_p|}$ by concatenating its title and abstract. The metadata of a paper $p$ is represented as a set $\mathcal{M}_p=\{m_1,m_2,..,m_{|\mathcal{M}_p|}\}$ consisting of its venue, author(s), and reference(s).


\vspace{1mm}

\noindent \textbf{Problem Definition.} The scientific literature tagging task can be cast as a large-scale multi-label classification problem, where all candidate tags (e.g., ``\textsf{Organic Chemistry}'', ``\textsf{Coupling Reaction}'', ``\textsf{Suzuki Reaction}'') constitute a large label space $\mathcal{L}$ (e.g., with $10^3$--$10^5$ labels). Given a scientific paper with its text and metadata information, the task is to find a set of labels from $\mathcal{L}$ that are relevant to the paper. Formally, the problem is defined as follows.

\begin{definition}{(Problem Definition)}
Given a training corpus $\mathcal{D}$ and the label space $\mathcal{L}$, where each paper $p \in \mathcal{D}$ is associated with its text $\mathcal{T}_p$, metadata information $\mathcal{M}_p$, and relevant tags $\mathcal{L}_p \subseteq \mathcal{L}$, our objective is to learn a multi-label text classifier $f_{\rm class}$ that can map an unlabeled paper $p' \notin \mathcal{D}$ to its relevant tags $\mathcal{L}_{p'} \subseteq \mathcal{L}$.
\label{def:problem}
\end{definition}

\section{Dataset Construction}
\label{sec:data}
\begin{table}[t]
\centering
\caption{Statistics of the 20 datasets in \textsc{\dataset} across 19 fields. There are 2 datasets in the Computer Science field, one of which is collected from top conferences and the other from top journals.}
\vspace{-0.5em}
\scalebox{0.68}{
\begin{NiceTabular}{c|cccccc}
\hline
\textbf{Field} 
& \textbf{\begin{tabular}[c]{@{}c@{}}Paper\\ Source\end{tabular}} 
& \textbf{\#Papers} 
& \textbf{\#Labels}  
& \textbf{\#Venues} 
& \textbf{\#Authors}
& \textbf{\#References}
\\ \hline 
\rowcolor{black!15}
Art                             & Journal         & 58,373            & 1,990             & 98                & 54,802             & 115,343            \\
Philosophy                      & Journal         & 59,296            & 3,758             & 98                & 36,619             & 198,010            \\
\rowcolor{black!15}
Geography                       & Journal         & 73,883            & 3,285             & 98                & 157,423            & 884,632            \\
Business                        & Journal         & 84,858            & 2,392             & 97                & 100,525            & 685,034            \\
\rowcolor{black!15}
Sociology                       & Journal         & 90,208            & 1,935             & 98                & 85,793             & 842,561            \\
History                         & Journal         & 113,147           & 2,689             & 99                & 84,529             & 284,739            \\
\rowcolor{black!15}
\makecell{Political \\ Science}                   & Journal         & 115,291           & 4,990             & 98                & 93,393             & 480,136             \\
\makecell{Environmental \\ Science}               & Journal         & 123,945           & 694               & 100               & 265,728            & 1,217,268           \\
\rowcolor{black!15}
Economics                       & Journal         & 178,670           & 5,205             & 97                & 135,247            & 1,042,253           \\
Engineering                     & Journal         & 270,006           & 10,683            & 100               & 430,046            & 1,867,276           \\
\rowcolor{black!15}
Psychology                      & Journal         & 372,954           & 7,641             & 100               & 460,123            & 2,313,701           \\
\multirow{2}{*}{\begin{tabular}[c]{@{}c@{}}Computer \\ Science\end{tabular}}   & Conference      & 263,393           & 13,613            & 75                & 331,582            & 1,084,440           \\
                                & Journal         & 410,603           & 15,540            & 96                & 634,506            & 2,751,996           \\
\rowcolor{black!15}
Geology                         & Journal         & 431,834           & 7,883             & 100               & 471,216            & 1,753,762           \\
Mathematics                     & Journal         & 490,551           & 14,271            & 98                & 404,066            & 2,150,584           \\
\rowcolor{black!15}
\makecell{Materials \\ Science}                   & Journal         & 1,337,731         & 6,802             & 99                & 1,904,549          & 5,457,773           \\
Physics                         & Journal         & 1,369,983         & 16,664            & 91                & 1,392,070          & 3,641,761           \\
\rowcolor{black!15}
Biology                         & Journal         & 1,588,778         & 64,267            & 100               & 2,730,547          & 7,086,131           \\
Chemistry                       & Journal         & 1,849,956         & 35,538            & 100               & 2,721,253          & 8,637,438           \\
\rowcolor{black!15}
Medicine                        & Journal         & 2,646,105         & 36,619            & 100               & 4,345,385          & 7,405,779           \\ \hline
\end{NiceTabular}
}
\vspace{-1em}
\label{tab:dataset}
\end{table}

Previous studies on scientific literature tagging \cite{peng2016deepmesh,xun2019meshprobenet,you2021bertmesh,dai2020fullmesh,zhang2021match,ye2021beyond} mainly use computer science and biomedicine papers to evaluate their proposed model, meanwhile paying less attention to other scientific fields. To bridge this gap, we construct \textsc{\dataset}, a multi-field benchmark for evaluating scientific literature tagging. \textsc{\dataset} is built upon data from the Microsoft Academic Graph (MAG) \cite{sinha2015overview}, which 
has been widely adopted in scientific text mining \cite{cohan2020specter,zhang2021match,yin2022public}. MAG covers 19 academic fields, which are listed in Table \ref{tab:dataset}. For each field, we conduct the following steps to construct a dataset.

\vspace{1mm}

\noindent \textbf{Venue Selection.} MAG maintains a list of the top-100 journals in each field according to the $h$-index \cite{hirsch2005index}. When constructing \textsc{\dataset}, we focus on papers published in these top journals. Note that some preprint services (e.g., ``\textit{arxiv}'', ``\textit{bioRxiv}'', ``\textit{SSRN}'', ``\textit{NBER Working Paper}'') \cite{xie2021preprint} are also viewed as top journals in MAG, but we exclude them from our consideration. As a result, in Table \ref{tab:dataset}, some of the constructed datasets contain less than 100 venues. Among the 19 fields, computer science (CS) has a unique publication culture: CS papers often appear first or exclusively in conferences rather than journals. Thus, for the Computer Science field, besides a collection of journal papers, we construct another dataset with papers from 75 top conferences according to CSRankings\footnote{\url{https://csrankings.org/}}. That being said, we will construct 20 datasets in total for the 19 fields.

\vspace{1mm}

\noindent \textbf{Label Space Construction.} For each field, we need a set of candidate labels for paper tagging. MAG has a directed acyclic graph (DAG)-structured label taxonomy $\mathcal{L}_{\rm MAG}$ \cite{shen2018web}. The taxonomy has 6 levels and more than $10^5$ labels, where each of the 19 fields is a Layer-0 label (i.e., the most coarse-grained label). Given a field $F$, we extract its descendant labels in the taxonomy $\mathcal{L}_{\rm MAG}$ as the label space $\mathcal{L}_F$ of this field, but we exclude $F$ itself from $\mathcal{L}_F$ because the root label is trivial to predict in classification. Since a label may have more than one parent in the DAG-structured taxonomy, it may appear in the label space of two or more fields. For example, the label ``\textsf{Anatomy}'' is a child of both ``\textsf{Biology}'' and ``\textsf{Medicine}'', so it is a candidate label in both the Biology and the Medicine datasets in \textsc{\dataset}.

\vspace{1mm}

\noindent \textbf{Paper Selection.} Each paper $p$ in MAG is tagged with its relevant labels $\mathcal{L}_p \subseteq \mathcal{L}_{\rm MAG}$ \cite{shen2018web}\footnote{The paper tags $\mathcal{L}_p$ come from the predictions of a system \cite{shen2018web} proposed by Microsoft Academic. The tags are accurate as checked by humans \cite{shen2018web} and have been used to support notable findings \cite{yin2022public,jin2021scientific}. Meanwhile, we also conduct experiments on three datasets with MeSH labels \cite{coletti2001medical}, which are curated by biomedical experts. Discussions on the additional three datasets can be found in Appendix \ref{sec:app_mesh}.}. To be included in \textsc{\dataset} (given a field $F$), a paper $p$ needs to satisfy the following two criteria: (1) $p$ is published in a selected venue of $F$; (2) $p$ is labeled with $F$ and at least one of $F$'s descendants (i.e., $F \in \mathcal{L}_p$ and $|\mathcal{L}_p \cap \mathcal{L}_F| \geq 1$). When studying the scientific literature tagging task in a field $F$, we focus on labels related to that field only. Therefore, the ground truth labels of $p$ are defined as $\mathcal{L}_{p|F} = \mathcal{L}_p \cap \mathcal{L}_F$. Note that a paper may appear in more than one field, and its ground truth labels are different when we consider different fields. For example, if a paper is tagged with ``\textsf{Medicine}'', ``\textsf{Polypharmacy}'', ``\textsf{Computer Science}'', and ``\textsf{Graph Embedding}'', given that ``\textsf{Polypharmacy}'' is a candidate label in the Medicine field and ``\textsf{Graph Embedding}'' is a candidate label in the Computer Science field, the paper will appear in both the Medicine and the Computer Science datasets in \textsc{\dataset}. However, when we perform tagging in the Medicine field, the ground-truth label of the paper is ``\textsf{Polypharmacy}''; when we perform tagging in the Computer Science field, the ground-truth label of the paper is ``\textsf{Graph Embedding}''.

\vspace{1mm}

\noindent \textbf{Text and Metadata Extraction.} For each selected paper $p$, we extract its title, abstract, venue, author(s), and reference(s) from MAG. The title and abstract are concatenated as text information $\mathcal{T}_p$. The venue, author(s), and reference(s) constitute metadata features $\mathcal{M}_p$ of the paper.

\vspace{1mm}

\noindent \textbf{Training-Validation-Testing Split.} \textsc{\dataset} contains academic papers published between Jan. 1, 1981 and Dec. 31, 2020. We use papers from 1981 to 2015 for training and validation, and papers from 2016 to 2020 for testing.

Statistics of the constructed 20 datasets in \textsc{\dataset} can be found in Table \ref{tab:dataset}. More details are shown in Appendix \ref{sec:app_data}. From now on, for convenience of discussion, we use the terms ``field'' and ``dataset'' interchangeably if there is no ambiguity. (In other words, we treat ``Computer Science (Conference)'' and ``Computer Science (Journal)'' as different fields.)

\section{Models}
\label{sec:model}
Large-scale multi-label text classification (LMTC) has been extensively studied over the past decade. Various approaches are proposed and can be applied to scientific literature tagging. Based on how text is used as features in the classifier, existing LMTC approaches can be divided into three major categories: (1) \textit{Bag-of-words classifiers} \cite{prabhu2018parabel,babbar2017dismec,yen2017ppdsparse,prabhu2018extreme,jain2016extreme,prabhu2014fastxml,tagami2017annexml,yen2016pd,peng2016deepmesh,yu2022pecos,khandagale2020bonsai} treat each document as a multiset of tokens while disregarding word position and order. Trees, embeddings, and linear layers are commonly used in these models. (2) \textit{Sequence-based classifiers} \cite{liu2017deep,xun2019meshprobenet,you2019attentionxml,zhang2021match,ye2021beyond,xun2020correlation} take each document as a sequence of tokens and train a CNN, RNN, or Transformer architecture from scratch to build a multi-label classifier. (3) \textit{Pre-trained language model classifiers} \cite{chang2020taming,you2021bertmesh,zhang2022metadata,jiang2021lightxml,ye2020pretrained,zhang2021fast} aim at transferring the knowledge learned from web-scale corpora (e.g., Wikipedia and PubMed) to the LMTC task, which can complement the text information from the training corpus. Since the goal of this paper is to study the effect of metadata, we select one LMTC model from each of the three categories that can be easily augmented with metadata information. Now we introduce the three selected models -- Parabel \cite{prabhu2018parabel} (bag-of-words), Transformer \cite{xun2020correlation} (sequence-based), and OAG-BERT \cite{liu2022oag} (pre-trained language model) -- and how they can take text and metadata features as input for classification.

\subsection{Bag-of-Words Classifier: Parabel {\small \cite{prabhu2018parabel}}}
\subsubsection{Using Text Only} 
In general, bag-of-words classifiers represent each document $p$ as a $|\mathcal{V}_{\mathcal{D}}|$-dimensional vector $\bmx_p$, where $\mathcal{V}_{\mathcal{D}}$ is the vocabulary of the training corpus $\mathcal{D}$. Given a word $w \in \mathcal{V}_{\mathcal{D}}$, its corresponding entry in $\bmx_p$ is defined using the tf--idf score:
\begin{equation}
\small
    x_{p, w} = {\rm tf}(w, p) \cdot {\rm idf}(w, \mathcal{D}).
\end{equation}
Here, ${\rm tf}(w, p)$ is the term frequency of $w$ in document $p$; ${\rm idf}(w, \mathcal{D}) = \log\frac{|\mathcal{D}|}{|\{p'\in \mathcal{D}:\ w\in \mathcal{T}_{p'}\}|}$ is the inverse document frequency of $w$.

The relevant labels of each document are represented by an $|\mathcal{L}|$-dimensional vector $\bmy_p$, where $y_{p,l}=1$ if $l$ is a tag relevant to $p$ (i.e., $l \in \mathcal{L}_p$), and $y_{p,l}=0$ otherwise.


To perform multi-label text classification, Parabel learns an ensemble of multiple label trees, each of which is obtained by recursively partitioning the labels into two balanced groups until each node contains less than a certain number of labels. The partition process is implemented by spherical 2-means clustering based on the label representation $\bmv_l$, which is a unit vector in the direction of the mean of the training points containing label $l$. After label space partitioning, Parabel learns a hierarchical discriminative classifier $\Pr(\bmy_p|\bmx_p)$. Specifically, at each non-leaf node, a distribution is learned to determine which child nodes should be traversed; at each leaf node, a distribution is learned to predict the set of relevant tags. For more technical details, one can refer to \cite{prabhu2018parabel}, but we omit them here as they are not directly related to the usage of metadata.

\subsubsection{Using Text + Metadata}
It is straightforward to generalize bag-of-words classifiers to take metadata features. Let $\mathcal{U}_{\mathcal{D}}$ denote the set of metadata instances appearing in $\mathcal{D}$. Given a metadata instance $m \in \mathcal{U}_{\mathcal{D}}$ and a paper $p$, we define ${\rm tf}(m, p)$ and ${\rm idf}(m, \mathcal{D})$ as follows:
\begin{equation}
\small
    {\rm tf}(m, p) = {\bf 1}(m \in \mathcal{M}_p), \ \ \ {\rm idf}(m, \mathcal{D}) = \log\frac{|\mathcal{D}|}{|\{p'\in \mathcal{D}: m \in \mathcal{M}_{p'}\}|},
\label{eqn:meta_tf}
\end{equation}
where ${\bf 1}(\cdot)$ is the indicator function. One can find that the definitions in Eq. (\ref{eqn:meta_tf}) are well aligned with the definitions of ${\rm tf}(w, p)$ and ${\rm idf}(w, \mathcal{D})$, except that a metadata instance does not appear multiple times in a document.

Now we can have a ``bag-of-metadata'' representation $\tilde{\bmx}_p$ for each paper. $\tilde{\bmx}_p$ is a $|\mathcal{U}_{\mathcal{D}}|$-dimensional vector, where
\begin{equation}
\small
    \tilde{x}_{p, m} = {\rm tf}(m, p) \cdot {\rm idf}(m, \mathcal{D}).
\end{equation}
Finally, we represent each paper $p$ as a $(|\mathcal{U}_{\mathcal{D}}|+|\mathcal{V}_{\mathcal{D}}|)$-dimensional vector, that is, the concatenation of bag-of-words and ``bag-of-metadata'' representations $\bmx_p \ || \ \tilde{\bmx}_p$. This vector is fed into Parabel for training and prediction.

\subsection{Sequence-based Classifier: Transformer {\small \cite{xun2020correlation}}}
\subsubsection{Using Text Only} 
To apply Transformer \cite{vaswani2017attention} to LMTC, we follow the architecture proposed in \cite{xun2020correlation}, which adds multiple ``[CLS]'' tokens in front of document text $\mathcal{T}_p$ as the input sequence. Formally, given $\mathcal{T}_p=w_1w_2...w_{|\mathcal{T}_p|}$, the input sequence of paper $p$ is
\begin{equation}
\small
\mathcal{I}_p = [{\rm CLS}_1] \ [{\rm CLS}_2] \ ... \ [{\rm CLS}_C] \ w_1 \ w_2 \ ... \ w_{|\mathcal{T}_p|}
\label{eqn:text_input}
\end{equation}
The motivation here is that when the label space is large (e.g., with $10^4$ tags), the output representation of one ``[CLS]'' token (e.g., a vector with several hundred dimensions) may not carry enough information to predict relevant labels. Therefore, multiple ``[CLS]'' tokens are needed to probe the semantics of text from different perspectives.

After Transformer encodes the input sequence $\mathcal{I}_p$, each token $w \in \mathcal{I}_p$ will have an output representation $\bmh_w$. We concatenate the representations of all ``[CLS]'' tokens together as the paper embedding:
\begin{equation}
\small
\bmh_p = \bmh_{[{\rm CLS}_1]} \ || \ \bmh_{[{\rm CLS}_2]} \ || \ ... \ || \ \bmh_{[{\rm CLS}_C]}.
\label{eqn:cls}
\end{equation}
$\bmh_p$ is further fed into a fully connected layer to perform multi-label classification:
\begin{equation}
\small
\hat{\bmy}_p = {\rm Sigmoid}({\bm W}^{\top} \bmh_p + {\bm b}).
\end{equation}
Here, $\hat{\bmy}_p$ is an $|\mathcal{L}|$-dimensional vector, in which $\hat{y}_{p,l}$ is the predicted probability that paper $p$ is relevant to label $l$. The classifier is trained to minimize the following binary cross-entropy (BCE):
\begin{equation}
\small
-\sum_{l \in \mathcal{L}}(y_{p,l}\log\hat{y}_{p,l} + (1-y_{p,l})\log(1-\hat{y}_{p,l})).
\label{eqn:loss}
\end{equation}

\subsubsection{Using Text + Metadata} 
The fully connected attention mechanism in Transformer paves an easy way to incorporate metadata. To be specific, following \cite{zhang2021match}, we can directly insert metadata tokens into the input sequence. Given paper text $\mathcal{T}_p=w_1...w_{|\mathcal{T}_p|}$ and metadata $\mathcal{M}_p=\{m_1,...,m_{|\mathcal{M}_p|}\}$, the input sequence is
\begin{equation}
\small
\widetilde{\mathcal{I}}_p = [{\rm CLS}_1] \ ... \ [{\rm CLS}_C] \ m_1 \ ... \ m_{|\mathcal{M}_p|} \ [{\rm SEP}] \ w_1 \ ... \ w_{|\mathcal{T}_p|}
\end{equation}
Unlike in CNN \cite{liu2017deep} or RNN \cite{xun2019meshprobenet,you2019attentionxml} classifiers, in Transformer, the order of metadata instances does not matter much because each pair of (metadata, word) or (metadata, metadata) can interact with each other via the fully connected attention mechanism during encoding. In our experiments, when listing authors in $\widetilde{\mathcal{I}}_p$, we follow the authorship order (i.e., 1$^{\rm st}$ author, 2$^{\rm nd}$ author, ...); when listing references in $\widetilde{\mathcal{I}}_p$, we adopt a random order because the citation order and inline contexts are not stored in MAG.

Following \cite{zhang2021match}, we treat each metadata instance $m_i \in \widetilde{\mathcal{I}}_p$ as one token during Transformer encoding. For example, the venue ``\textit{WWW}'' is represented as one token ``[\textsc{Venue}\_1135342153]'' instead of its textual name ``\textit{the web conference}'' containing multiple tokens. 

After $\widetilde{\mathcal{I}}_p$ is fed to Transformer, the remaining designs of the metadata-aware classifier exactly follow Eqs. (\ref{eqn:cls})-(\ref{eqn:loss}).

\subsection{Pre-trained Language Model Classifier: OAG-BERT {\small \cite{liu2022oag}}}
\begin{figure}[t]
\centering
\subfloat[\rm The pre-training process of OAG-BERT \cite{liu2022oag}]{
\includegraphics[width=0.9\linewidth]{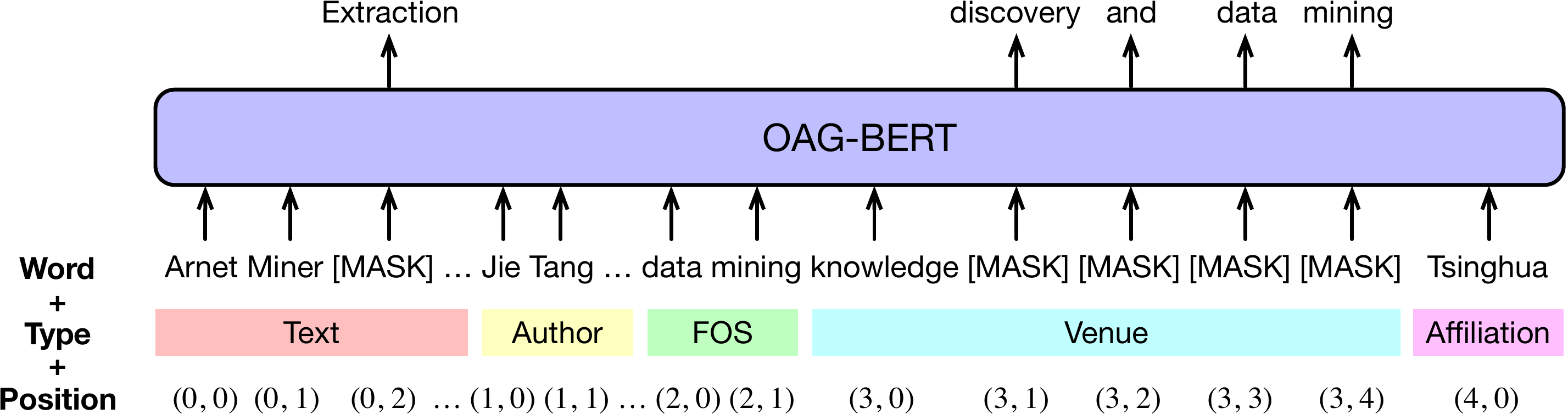}} \\
\vspace{-0.8em}
\subfloat[\rm Encoding a paper when we use text only]{
\includegraphics[width=0.9\linewidth]{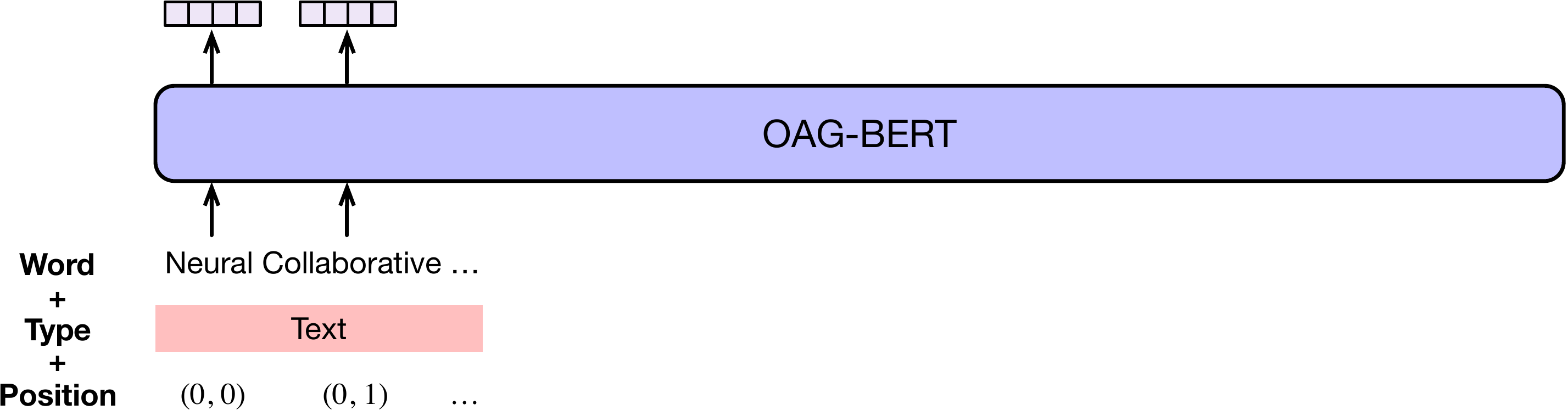}} \\
\vspace{-0.8em}
\subfloat[\rm Encoding a paper when we use text + metadata]{
\includegraphics[width=0.9\linewidth]{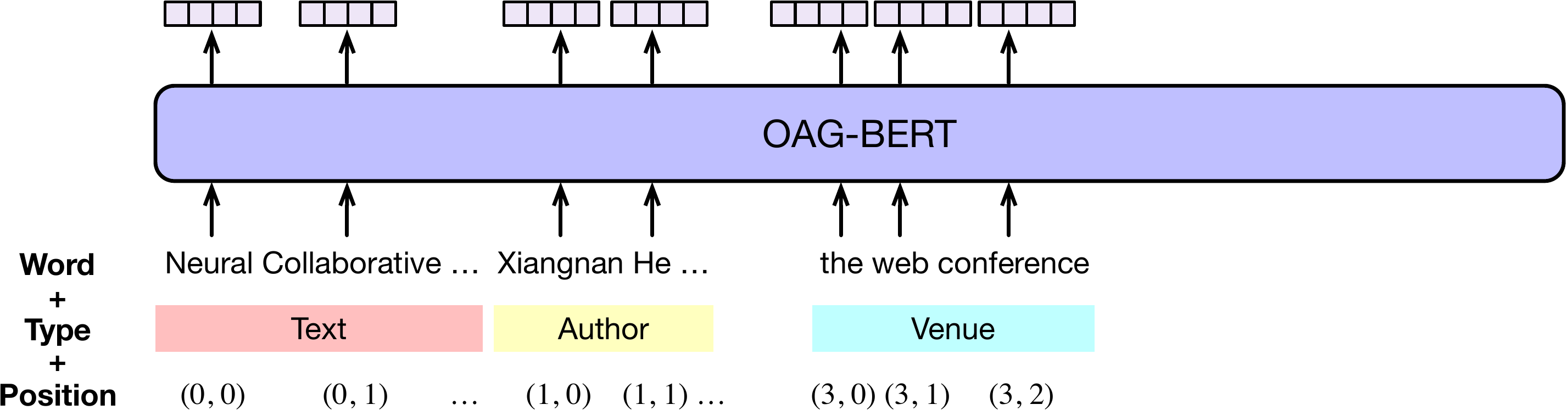}} \\
\vspace{-0.5em}
\caption{The pre-training process and our usage of OAG-BERT \cite{liu2022oag}.}
\vspace{-1em}
\label{fig:oagbert}
\end{figure}

Previous classifiers built upon pre-trained language models (PLMs) mainly utilize BERT \cite{devlin2019bert}, BioBERT \cite{lee2020biobert}, SciBERT \cite{beltagy2019scibert}, XLNet \cite{yang2019xlnet}, or RoBERTa \cite{liu2019roberta} to derive document representations. However, these PLMs mostly focus on text information during pre-training and are not specifically designed to deal with metadata. To bridge this gap, we adopt OAG-BERT \cite{liu2022oag}, an entity-augmented academic language model, which can jointly encode scientific text and venue/author information.

The pre-training process of OAG-BERT is briefly illustrated in Figure \ref{fig:oagbert}(a). It places text and metadata information in a single sequence for masked language modeling. Besides, it proposes three strategies to deal with metadata entities: (1) heterogeneous entity type embedding makes the model aware of different metadata types; (2) span-aware entity masking selects a continuous span within long entities (e.g., the venue ``\textit{knowledge discovery and data mining}''); (3) 2-dimensional positional
embedding jointly models inter and intra-entity token orders. The model is first trained on 5 million full-text papers and then on 120 million paper titles/abstracts with metadata from the Open Academic Graph \cite{zhang2019oag}.

\subsubsection{Using Text Only}
When considering paper text $\mathcal{T}_p$ only for classification, we first use OAG-BERT to encode the text sequence (as illustrated in Figure \ref{fig:oagbert}(b)).
\begin{equation}
\small
\bmH_p = \text{OAG-BERT}(\mathcal{T}_p).    
\end{equation}
Here, $\bmH_p=[\bmh_{p,1},...,\bmh_{p,N}]$ contains the output vectors of all tokens. We then adopt mean pooling to obtain the paper representation $\bmx_p$.
\begin{equation}
\small
\bmx_p = \frac{1}{N}\sum_{i=1}^N \bmh_{p,i}.
\end{equation}

Directly fine-tuning the PLM using a BCE loss (i.e., Eq. (\ref{eqn:loss})) is very time-consuming, especially when the label space is large. Considering efficiency and model simplicity, we fix the paper representation $\bmx_p$ to train a Parabel classifier $f_{\rm Parabel}(\bmx_p)=\Pr(\bmy_p|\bmx_p)$.

During the prediction stage, given a paper $p'$, the trained classifier predicts the probability that $p'$ is relevant to each label $l$:
\begin{equation}
\small
    \hat{\bmy}_{p'} = f_{\rm Parabel}(\bmx_{p'}).
\end{equation}Then, we can rank the labels according to $\hat{y}_{p',l}$ to predict a list of the most relevant labels. However, in practice, we find this strategy does not yield competitive prediction accuracy. This finding is consistent with the design in related studies \cite{chang2020taming,peng2016deepmesh} that discrete lexical features should be considered together with continuous representations during classification. To add lexical features, motivated by \cite{zhang2022metadata}, we propose a simple heuristic to re-rank the labels: Given a paper $p'$, all labels whose name appears in the paper text $\mathcal{T}_{p'}$ will be ranked higher than those not appearing in $\mathcal{T}_{p'}$. This heuristic is equivalent to ranking the labels according to a modified score $\hat{z}_{p',l} = \hat{y}_{p',l}+{\bf 1}(t_l \subseteq \mathcal{T}_{p'})$, where $t_l$ stands for the textual name of label $l$.

\subsubsection{Using Text + Metadata}
When metadata information is available, we use OAG-BERT to jointly encode paper text and metadata (as illustrated in Figure \ref{fig:oagbert}(c)).
\begin{equation}
\small
\widetilde{\bmH}_p = \text{OAG-BERT}(\mathcal{T}_p, \mathcal{M}_p).    
\end{equation}
The remaining steps exactly follow the text-only model. It is worth noting that since references are not involved during the pre-training of OAG-BERT, we can only use venues and authors as metadata here. Also, unlike the Transformer classifier that views each venue/author as one token, OAG-BERT tokenizes the textual name of venues/ authors based on its vocabulary during encoding.

\begin{table*}[!h]
\centering
\renewcommand\arraystretch{0.9}
\caption{P@$k$ and NDCG@$k$ scores on the 20 datasets. \colorbox{blue!15}{Blue}: significantly better than using text only (p-value $< 0.05$). \colorbox{red!15}{Red}: significantly worse than using text only (p-value $< 0.05$). ``--'': OAG-BERT cannot take references as metadata signals.}
\vspace{-0.2em}
\scalebox{0.72}{
\begin{NiceTabular}{c|c|ccccc|ccccc|ccccc}[colortbl-like]
\hline
\multirow{2}{*}{\textbf{Field}} & \multirow{2}{*}{\textbf{Input}} & \multicolumn{5}{c|}{\textbf{Parabel \cite{prabhu2018parabel}}} & \multicolumn{5}{c|}{\textbf{Transformer \cite{xun2020correlation}}} & \multicolumn{5}{c}{\textbf{OAG-BERT \cite{liu2022oag}}} \\ \cline{3-17}
 & & \textbf{P@1} & \textbf{P@3} & \textbf{P@5} & \textbf{N@3} & \textbf{N@5} & \textbf{P@1} & \textbf{P@3} & \textbf{P@5} & \textbf{N@3} & \textbf{N@5} & \textbf{P@1} & \textbf{P@3} & \textbf{P@5} & \textbf{N@3} & \textbf{N@5} \\ \hline
& Text & 0.7203 & 0.4829 & 0.3392 & 0.7585 & 0.7921 & 0.6995 & 0.4396 & 0.3008 & 0.6956 & 0.7183 & 0.6982 & 0.4411 & 0.3030 & 0.7024 & 0.7253 \\ 
& +Venue & {\cellcolor{blue!15} 0.7235} & {\cellcolor{blue!15} 0.4861} & {\cellcolor{blue!15} 0.3417} & {\cellcolor{blue!15} 0.7630} & {\cellcolor{blue!15} 0.7969} & {\cellcolor{blue!15} 0.7109} & {\cellcolor{blue!15} 0.4456} & {\cellcolor{blue!15} 0.3038} & {\cellcolor{blue!15} 0.7065} & {\cellcolor{blue!15} 0.7279} & 0.7032 & {\cellcolor{blue!15} 0.4476} & {\cellcolor{blue!15} 0.3066} & {\cellcolor{blue!15} 0.7116} & {\cellcolor{blue!15} 0.7335} \\ 
& +Author & {\cellcolor{blue!15} 0.7247} & 0.4827 & 0.3393 & 0.7598 & 0.7936 & 0.7060 & {\cellcolor{blue!15} 0.4447} & {\cellcolor{blue!15} 0.3035} & {\cellcolor{blue!15} 0.7024} & {\cellcolor{blue!15} 0.7238} & {\cellcolor{blue!15} 0.7032} & {\cellcolor{blue!15} 0.4454} & {\cellcolor{blue!15} 0.3062} & {\cellcolor{blue!15} 0.7085} & {\cellcolor{blue!15} 0.7318} \\ 
\multirow{-4}{*}{Art} & +Reference & 0.7186 & 0.4818 & 0.3389 & 0.7570 & 0.7913 & 0.7052 & {\cellcolor{blue!15} 0.4432} & {\cellcolor{blue!15} 0.3034} & {\cellcolor{blue!15} 0.6997} & {\cellcolor{blue!15} 0.7222} & -- & -- & -- & -- & -- \\ \hline 
& Text & 0.8124 & 0.6033 & 0.4617 & 0.7578 & 0.7434 & 0.7493 & 0.5103 & 0.3712 & 0.6524 & 0.6219 & 0.7652 & 0.5548 & 0.4157 & 0.6980 & 0.6738 \\ 
& +Venue & {\cellcolor{blue!15} 0.8164} & {\cellcolor{blue!15} 0.6322} & {\cellcolor{blue!15} 0.4909} & {\cellcolor{blue!15} 0.7834} & {\cellcolor{blue!15} 0.7738} & {\cellcolor{blue!15} 0.7889} & {\cellcolor{blue!15} 0.5725} & {\cellcolor{blue!15} 0.4285} & {\cellcolor{blue!15} 0.7159} & {\cellcolor{blue!15} 0.6897} & {\cellcolor{blue!15} 0.7679} & {\cellcolor{blue!15} 0.5621} & {\cellcolor{blue!15} 0.4271} & {\cellcolor{blue!15} 0.7045} & {\cellcolor{blue!15} 0.6852} \\ 
& +Author & 0.8128 & 0.6050 & {\cellcolor{blue!15} 0.4637} & 0.7593 & {\cellcolor{blue!15} 0.7455} & 0.7544 & 0.5151 & 0.3743 & {\cellcolor{blue!15} 0.6577} & {\cellcolor{blue!15} 0.6262} & 0.7673 & {\cellcolor{blue!15} 0.5594} & {\cellcolor{blue!15} 0.4206} & {\cellcolor{blue!15} 0.7030} & {\cellcolor{blue!15} 0.6799} \\ 
\multirow{-4}{*}{Philosophy} & +Reference & {\cellcolor{red!15} 0.8023} & {\cellcolor{red!15} 0.5968} & {\cellcolor{red!15} 0.4581} & {\cellcolor{red!15} 0.7478} & {\cellcolor{red!15} 0.7344} & {\cellcolor{red!15} 0.7442} & 0.5072 & 0.3698 & {\cellcolor{red!15} 0.6463} & {\cellcolor{red!15} 0.6160} & -- & -- & -- & -- & -- \\ \hline 
& Text & 0.6233 & 0.4114 & 0.3015 & 0.6331 & 0.6640 & 0.5239 & 0.3140 & 0.2237 & 0.4887 & 0.5036 & 0.6220 & 0.3997 & 0.2843 & 0.6158 & 0.6341 \\ 
& +Venue & {\cellcolor{blue!15} 0.6283} & {\cellcolor{blue!15} 0.4147} & {\cellcolor{blue!15} 0.3036} & {\cellcolor{blue!15} 0.6384} & {\cellcolor{blue!15} 0.6694} & {\cellcolor{blue!15} 0.5457} & {\cellcolor{blue!15} 0.3251} & {\cellcolor{blue!15} 0.2304} & {\cellcolor{blue!15} 0.5083} & {\cellcolor{blue!15} 0.5223} & 0.6244 & {\cellcolor{blue!15} 0.4036} & {\cellcolor{blue!15} 0.2879} & {\cellcolor{blue!15} 0.6213} & {\cellcolor{blue!15} 0.6406} \\ 
& +Author & 0.6215 & 0.4113 & {\cellcolor{blue!15} 0.3019} & 0.6321 & 0.6637 & 0.5241 & 0.3141 & 0.2239 & 0.4888 & 0.5038 & 0.6237 & {\cellcolor{blue!15} 0.4046} & {\cellcolor{blue!15} 0.2885} & {\cellcolor{blue!15} 0.6219} & {\cellcolor{blue!15} 0.6414} \\ 
\multirow{-4}{*}{Geography} & +Reference & {\cellcolor{red!15} 0.6135} & {\cellcolor{red!15} 0.4104} & {\cellcolor{blue!15} 0.3021} & {\cellcolor{red!15} 0.6277} & {\cellcolor{red!15} 0.6603} & {\cellcolor{red!15} 0.4960} & {\cellcolor{red!15} 0.3030} & {\cellcolor{red!15} 0.2181} & {\cellcolor{red!15} 0.4665} & {\cellcolor{red!15} 0.4835} & -- & -- & -- & -- & -- \\ \hline 
& Text & 0.6601 & 0.4775 & 0.3666 & 0.6192 & 0.6351 & 0.6375 & 0.4344 & 0.3238 & 0.5736 & 0.5780 & 0.6687 & 0.4859 & 0.3648 & 0.6328 & 0.6408 \\ 
& +Venue & {\cellcolor{blue!15} 0.6667} & {\cellcolor{blue!15} 0.4811} & {\cellcolor{blue!15} 0.3690} & {\cellcolor{blue!15} 0.6247} & {\cellcolor{blue!15} 0.6402} & {\cellcolor{blue!15} 0.6550} & {\cellcolor{blue!15} 0.4435} & {\cellcolor{blue!15} 0.3293} & {\cellcolor{blue!15} 0.5878} & {\cellcolor{blue!15} 0.5914} & 0.6693 & {\cellcolor{blue!15} 0.4899} & {\cellcolor{blue!15} 0.3688} & {\cellcolor{blue!15} 0.6371} & {\cellcolor{blue!15} 0.6462} \\ 
& +Author & {\cellcolor{blue!15} 0.6626} & 0.4782 & {\cellcolor{blue!15} 0.3673} & {\cellcolor{blue!15} 0.6208} & {\cellcolor{blue!15} 0.6367} & {\cellcolor{red!15} 0.6265} & {\cellcolor{red!15} 0.4255} & {\cellcolor{red!15} 0.3161} & {\cellcolor{red!15} 0.5619} & {\cellcolor{red!15} 0.5649} & 0.6701 & {\cellcolor{blue!15} 0.4886} & {\cellcolor{blue!15} 0.3679} & {\cellcolor{blue!15} 0.6362} & {\cellcolor{blue!15} 0.6455} \\ 
\multirow{-4}{*}{Business} & +Reference & 0.6587 & {\cellcolor{blue!15} 0.4798} & {\cellcolor{blue!15} 0.3694} & {\cellcolor{blue!15} 0.6219} & {\cellcolor{blue!15} 0.6389} & {\cellcolor{red!15} 0.5872} & {\cellcolor{red!15} 0.3985} & {\cellcolor{red!15} 0.2947} & {\cellcolor{red!15} 0.5237} & {\cellcolor{red!15} 0.5242} & -- & -- & -- & -- & -- \\ \hline 
& Text & 0.6173 & 0.4207 & 0.3057 & 0.6219 & 0.6509 & 0.5851 & 0.3611 & 0.2514 & 0.5489 & 0.5614 & 0.6177 & 0.4004 & 0.2799 & 0.5998 & 0.6144 \\ 
& +Venue & {\cellcolor{blue!15} 0.6227} & {\cellcolor{blue!15} 0.4257} & {\cellcolor{blue!15} 0.3080} & {\cellcolor{blue!15} 0.6297} & {\cellcolor{blue!15} 0.6573} & {\cellcolor{blue!15} 0.6026} & {\cellcolor{blue!15} 0.3704} & {\cellcolor{blue!15} 0.2576} & {\cellcolor{blue!15} 0.5644} & {\cellcolor{blue!15} 0.5768} & 0.6195 & {\cellcolor{blue!15} 0.4046} & {\cellcolor{blue!15} 0.2834} & {\cellcolor{blue!15} 0.6048} & {\cellcolor{blue!15} 0.6202} \\ 
& +Author & 0.6182 & 0.4206 & 0.3054 & 0.6220 & 0.6505 & {\cellcolor{blue!15} 0.5970} & 0.3625 & 0.2517 & 0.5548 & 0.5664 & 0.6191 & {\cellcolor{blue!15} 0.4027} & {\cellcolor{blue!15} 0.2820} & 0.6026 & {\cellcolor{blue!15} 0.6179} \\ 
\multirow{-4}{*}{Sociology} & +Reference & {\cellcolor{red!15} 0.6112} & 0.4210 & 0.3068 & 0.6200 & 0.6500 & 0.5805 & {\cellcolor{red!15} 0.3573} & {\cellcolor{red!15} 0.2491} & {\cellcolor{red!15} 0.5395} & {\cellcolor{red!15} 0.5516} & -- & -- & -- & -- & -- \\ \hline 
& Text & 0.7164 & 0.4574 & 0.3173 & 0.7480 & 0.7799 & 0.6941 & 0.4210 & 0.2869 & 0.6928 & 0.7174 & 0.6591 & 0.4122 & 0.2822 & 0.6776 & 0.7029 \\ 
& +Venue & {\cellcolor{blue!15} 0.7199} & {\cellcolor{blue!15} 0.4615} & {\cellcolor{blue!15} 0.3200} & {\cellcolor{blue!15} 0.7537} & {\cellcolor{blue!15} 0.7856} & {\cellcolor{blue!15} 0.7008} & {\cellcolor{blue!15} 0.4264} & {\cellcolor{blue!15} 0.2897} & {\cellcolor{blue!15} 0.7010} & {\cellcolor{blue!15} 0.7245} & 0.6602 & {\cellcolor{blue!15} 0.4147} & {\cellcolor{blue!15} 0.2841} & {\cellcolor{blue!15} 0.6809} & {\cellcolor{blue!15} 0.7064} \\ 
& +Author & 0.7144 & 0.4571 & 0.3171 & 0.7469 & 0.7789 & 0.6988 & {\cellcolor{blue!15} 0.4233} & 0.2881 & 0.6959 & 0.7198 & 0.6627 & {\cellcolor{blue!15} 0.4161} & {\cellcolor{blue!15} 0.2853} & {\cellcolor{blue!15} 0.6832} & {\cellcolor{blue!15} 0.7089} \\ 
\multirow{-4}{*}{History} & +Reference & 0.7142 & 0.4565 & 0.3169 & 0.7463 & 0.7786 & {\cellcolor{blue!15} 0.6999} & {\cellcolor{blue!15} 0.4248} & {\cellcolor{blue!15} 0.2892} & {\cellcolor{blue!15} 0.6990} & {\cellcolor{blue!15} 0.7233} & -- & -- & -- & -- & -- \\ \hline 
& Text & 0.7723 & 0.5153 & 0.3767 & 0.7252 & 0.7291 & 0.7624 & 0.4863 & 0.3429 & 0.6892 & 0.6797 & 0.7327 & 0.4805 & 0.3393 & 0.6721 & 0.6624 \\ 
& +Venue & {\cellcolor{blue!15} 0.7776} & {\cellcolor{blue!15} 0.5216} & {\cellcolor{blue!15} 0.3817} & {\cellcolor{blue!15} 0.7327} & {\cellcolor{blue!15} 0.7374} & {\cellcolor{blue!15} 0.7723} & {\cellcolor{blue!15} 0.4948} & {\cellcolor{blue!15} 0.3494} & {\cellcolor{blue!15} 0.7004} & {\cellcolor{blue!15} 0.6910} & 0.7320 & {\cellcolor{blue!15} 0.4843} & {\cellcolor{blue!15} 0.3438} & 0.6756 & {\cellcolor{blue!15} 0.6672} \\ 
& +Author & 0.7737 & {\cellcolor{blue!15} 0.5173} & {\cellcolor{blue!15} 0.3779} & {\cellcolor{blue!15} 0.7275} & {\cellcolor{blue!15} 0.7313} & 0.7644 & 0.4882 & 0.3439 & 0.6913 & 0.6810 & {\cellcolor{blue!15} 0.7356} & {\cellcolor{blue!15} 0.4850} & {\cellcolor{blue!15} 0.3436} & {\cellcolor{blue!15} 0.6771} & {\cellcolor{blue!15} 0.6680} \\ 
\multirow{-4}{*}{\begin{tabular}[c]{@{}c@{}}Political\\ Science\end{tabular}} & +Reference & {\cellcolor{red!15} 0.7633} & {\cellcolor{red!15} 0.5107} & {\cellcolor{red!15} 0.3750} & {\cellcolor{red!15} 0.7182} & {\cellcolor{red!15} 0.7237} & {\cellcolor{red!15} 0.7549} & {\cellcolor{red!15} 0.4812} & {\cellcolor{red!15} 0.3392} & {\cellcolor{red!15} 0.6811} & {\cellcolor{red!15} 0.6712} & -- & -- & -- & -- & -- \\ \hline 
& Text & 0.6484 & 0.4254 & 0.3087 & 0.6990 & 0.7299 & 0.6447 & 0.4003 & 0.2856 & 0.6706 & 0.6964 & 0.7185 & 0.4465 & 0.3156 & 0.7449 & 0.7663 \\ 
& +Venue & 0.6488 & {\cellcolor{blue!15} 0.4269} & 0.3094 & 0.7005 & 0.7313 & 0.6469 & 0.4023 & 0.2867 & 0.6730 & 0.6981 & 0.7196 & 0.4468 & 0.3157 & 0.7454 & 0.7669 \\ 
& +Author & 0.6484 & {\cellcolor{blue!15} 0.4267} & 0.3095 & 0.7002 & 0.7313 & {\cellcolor{red!15} 0.6333} & {\cellcolor{red!15} 0.3945} & {\cellcolor{red!15} 0.2816} & {\cellcolor{red!15} 0.6590} & {\cellcolor{red!15} 0.6845} & 0.7189 & 0.4474 & {\cellcolor{blue!15} 0.3166} & 0.7461 & {\cellcolor{blue!15} 0.7683} \\ 
\multirow{-4}{*}{\begin{tabular}[c]{@{}c@{}}Environmental\\ Science\end{tabular}} & +Reference & 0.6486 & {\cellcolor{blue!15} 0.4278} & {\cellcolor{blue!15} 0.3112} & {\cellcolor{blue!15} 0.7016} & {\cellcolor{blue!15} 0.7340} & {\cellcolor{red!15} 0.6126} & {\cellcolor{red!15} 0.3843} & {\cellcolor{red!15} 0.2752} & {\cellcolor{red!15} 0.6388} & {\cellcolor{red!15} 0.6648} & -- & -- & -- & -- & -- \\ \hline 
& Text & 0.7288 & 0.5842 & 0.4793 & 0.6483 & 0.6301 & 0.7375 & 0.5785 & 0.4635 & 0.6448 & 0.6167 & 0.6861 & 0.5651 & 0.4612 & 0.6229 & 0.6037 \\ 
& +Venue & {\cellcolor{blue!15} 0.7339} & {\cellcolor{blue!15} 0.5885} & {\cellcolor{blue!15} 0.4830} & {\cellcolor{blue!15} 0.6534} & {\cellcolor{blue!15} 0.6353} & {\cellcolor{blue!15} 0.7460} & {\cellcolor{blue!15} 0.5848} & {\cellcolor{blue!15} 0.4690} & {\cellcolor{blue!15} 0.6521} & {\cellcolor{blue!15} 0.6243} & 0.6868 & {\cellcolor{blue!15} 0.5677} & {\cellcolor{blue!15} 0.4645} & {\cellcolor{blue!15} 0.6251} & {\cellcolor{blue!15} 0.6069} \\ 
& +Author & 0.7288 & {\cellcolor{blue!15} 0.5856} & {\cellcolor{blue!15} 0.4809} & {\cellcolor{blue!15} 0.6496} & {\cellcolor{blue!15} 0.6315} & {\cellcolor{red!15} 0.7224} & {\cellcolor{red!15} 0.5607} & {\cellcolor{red!15} 0.4477} & {\cellcolor{red!15} 0.6258} & {\cellcolor{red!15} 0.5967} & 0.6875 & {\cellcolor{blue!15} 0.5676} & {\cellcolor{blue!15} 0.4655} & {\cellcolor{blue!15} 0.6254} & {\cellcolor{blue!15} 0.6080} \\ 
\multirow{-4}{*}{Economics} & +Reference & 0.7306 & {\cellcolor{blue!15} 0.5866} & {\cellcolor{blue!15} 0.4826} & {\cellcolor{blue!15} 0.6511} & {\cellcolor{blue!15} 0.6340} & {\cellcolor{red!15} 0.6867} & {\cellcolor{red!15} 0.5303} & {\cellcolor{red!15} 0.4203} & {\cellcolor{red!15} 0.5921} & {\cellcolor{red!15} 0.5609} & -- & -- & -- & -- & -- \\ \hline 
& Text & 0.7158 & 0.5594 & 0.4597 & 0.6222 & 0.5909 & 0.6980 & 0.5287 & 0.4196 & 0.5924 & 0.5498 & 0.6516 & 0.5299 & 0.4349 & 0.5835 & 0.5541 \\ 
& +Venue & {\cellcolor{blue!15} 0.7196} & {\cellcolor{blue!15} 0.5638} & {\cellcolor{blue!15} 0.4638} & {\cellcolor{blue!15} 0.6268} & {\cellcolor{blue!15} 0.5958} & {\cellcolor{blue!15} 0.7079} & {\cellcolor{blue!15} 0.5400} & {\cellcolor{blue!15} 0.4300} & {\cellcolor{blue!15} 0.6040} & {\cellcolor{blue!15} 0.5618} & 0.6508 & 0.5308 & 0.4364 & 0.5832 & 0.5541 \\ 
& +Author & 0.7164 & {\cellcolor{blue!15} 0.5608} & {\cellcolor{blue!15} 0.4610} & {\cellcolor{blue!15} 0.6234} & {\cellcolor{blue!15} 0.5923} & {\cellcolor{red!15} 0.6770} & {\cellcolor{red!15} 0.5122} & {\cellcolor{red!15} 0.4068} & {\cellcolor{red!15} 0.5727} & {\cellcolor{red!15} 0.5308} & 0.6519 & 0.5315 & {\cellcolor{blue!15} 0.4375} & 0.5847 & 0.5561 \\ 
\multirow{-4}{*}{\begin{tabular}[c]{@{}c@{}}Computer\\ Science\\ (Conference)\end{tabular}} & +Reference & 0.7151 & {\cellcolor{blue!15} 0.5614} & {\cellcolor{blue!15} 0.4618} & 0.6236 & {\cellcolor{blue!15} 0.5927} & {\cellcolor{red!15} 0.6377} & {\cellcolor{red!15} 0.4777} & {\cellcolor{red!15} 0.3774} & {\cellcolor{red!15} 0.5353} & {\cellcolor{red!15} 0.4940} & -- & -- & -- & -- & -- \\ \hline 
& Text & 0.7881 & 0.6383 & 0.5274 & 0.7064 & 0.6776 & 0.7801 & 0.6154 & 0.4961 & 0.6852 & 0.6461 & 0.6780 & 0.5668 & 0.4757 & 0.6234 & 0.6058 \\ 
& +Venue & {\cellcolor{blue!15} 0.7904} & {\cellcolor{blue!15} 0.6399} & {\cellcolor{blue!15} 0.5285} & {\cellcolor{blue!15} 0.7087} & {\cellcolor{blue!15} 0.6798} & 0.7818 & 0.6168 & 0.4980 & 0.6869 & 0.6486 & {\cellcolor{blue!15} 0.6819} & {\cellcolor{blue!15} 0.5699} & {\cellcolor{blue!15} 0.4787} & {\cellcolor{blue!15} 0.6270} & {\cellcolor{blue!15} 0.6101} \\ 
& +Author & 0.7885 & {\cellcolor{blue!15} 0.6397} & {\cellcolor{blue!15} 0.5286} & {\cellcolor{blue!15} 0.7078} & {\cellcolor{blue!15} 0.6790} & {\cellcolor{red!15} 0.7679} & {\cellcolor{red!15} 0.5995} & {\cellcolor{red!15} 0.4820} & {\cellcolor{red!15} 0.6686} & {\cellcolor{red!15} 0.6286} & 0.6784 & {\cellcolor{blue!15} 0.5686} & {\cellcolor{blue!15} 0.4788} & {\cellcolor{blue!15} 0.6251} & {\cellcolor{blue!15} 0.6091} \\ 
\multirow{-4}{*}{Engineering} & +Reference & 0.7879 & {\cellcolor{blue!15} 0.6398} & {\cellcolor{blue!15} 0.5288} & {\cellcolor{blue!15} 0.7079} & {\cellcolor{blue!15} 0.6793} & {\cellcolor{red!15} 0.7261} & {\cellcolor{red!15} 0.5579} & {\cellcolor{red!15} 0.4453} & {\cellcolor{red!15} 0.6242} & {\cellcolor{red!15} 0.5832} & -- & -- & -- & -- & -- \\ \hline 
& Text & 0.7784 & 0.6272 & 0.5133 & 0.7073 & 0.6889 & 0.8035 & 0.6434 & 0.5172 & 0.7274 & 0.7007 & 0.7503 & 0.6148 & 0.4986 & 0.6909 & 0.6686 \\ 
& +Venue & {\cellcolor{blue!15} 0.7871} & {\cellcolor{blue!15} 0.6323} & {\cellcolor{blue!15} 0.5173} & {\cellcolor{blue!15} 0.7143} & {\cellcolor{blue!15} 0.6958} & {\cellcolor{blue!15} 0.8108} & {\cellcolor{blue!15} 0.6481} & {\cellcolor{blue!15} 0.5214} & {\cellcolor{blue!15} 0.7339} & {\cellcolor{blue!15} 0.7078} & {\cellcolor{blue!15} 0.7536} & {\cellcolor{blue!15} 0.6200} & {\cellcolor{blue!15} 0.5047} & {\cellcolor{blue!15} 0.6964} & {\cellcolor{blue!15} 0.6758} \\ 
& +Author & 0.7775 & {\cellcolor{blue!15} 0.6283} & {\cellcolor{blue!15} 0.5154} & 0.7078 & {\cellcolor{blue!15} 0.6903} & {\cellcolor{red!15} 0.7918} & {\cellcolor{red!15} 0.6300} & {\cellcolor{red!15} 0.5062} & {\cellcolor{red!15} 0.7126} & {\cellcolor{red!15} 0.6858} & 0.7489 & {\cellcolor{blue!15} 0.6165} & {\cellcolor{blue!15} 0.5015} & 0.6921 & {\cellcolor{blue!15} 0.6712} \\ 
\multirow{-4}{*}{Psychology} & +Reference & {\cellcolor{blue!15} 0.7821} & {\cellcolor{blue!15} 0.6334} & {\cellcolor{blue!15} 0.5209} & {\cellcolor{blue!15} 0.7136} & {\cellcolor{blue!15} 0.6973} & {\cellcolor{red!15} 0.7776} & {\cellcolor{red!15} 0.6205} & {\cellcolor{red!15} 0.4979} & {\cellcolor{red!15} 0.7017} & {\cellcolor{red!15} 0.6749} & -- & -- & -- & -- & -- \\ \hline 
& Text & 0.7075 & 0.5611 & 0.4645 & 0.6209 & 0.5862 & 0.7125 & 0.5477 & 0.4415 & 0.6110 & 0.5663 & 0.6356 & 0.5266 & 0.4416 & 0.5760 & 0.5481 \\ 
& +Venue & {\cellcolor{blue!15} 0.7098} & {\cellcolor{blue!15} 0.5639} & {\cellcolor{blue!15} 0.4673} & {\cellcolor{blue!15} 0.6238} & {\cellcolor{blue!15} 0.5893} & 0.7146 & {\cellcolor{blue!15} 0.5501} & {\cellcolor{blue!15} 0.4439} & {\cellcolor{blue!15} 0.6134} & {\cellcolor{blue!15} 0.5689} & 0.6359 & 0.5296 & {\cellcolor{blue!15} 0.4465} & 0.5789 & {\cellcolor{blue!15} 0.5529} \\ 
& +Author & {\cellcolor{blue!15} 0.7094} & {\cellcolor{blue!15} 0.5629} & {\cellcolor{blue!15} 0.4662} & {\cellcolor{blue!15} 0.6228} & {\cellcolor{blue!15} 0.5883} & {\cellcolor{red!15} 0.6937} & {\cellcolor{red!15} 0.5278} & {\cellcolor{red!15} 0.4241} & {\cellcolor{red!15} 0.5897} & {\cellcolor{red!15} 0.5449} & 0.6362 & 0.5299 & {\cellcolor{blue!15} 0.4467} & 0.5792 & {\cellcolor{blue!15} 0.5533} \\ 
\multirow{-4}{*}{\begin{tabular}[c]{@{}c@{}}Computer\\ Science\\ (Journal)\end{tabular}} & +Reference & {\cellcolor{blue!15} 0.7089} & {\cellcolor{blue!15} 0.5636} & {\cellcolor{blue!15} 0.4675} & {\cellcolor{blue!15} 0.6234} & {\cellcolor{blue!15} 0.5894} & {\cellcolor{red!15} 0.6534} & {\cellcolor{red!15} 0.4920} & {\cellcolor{red!15} 0.3936} & {\cellcolor{red!15} 0.5503} & {\cellcolor{red!15} 0.5062} & -- & -- & -- & -- & -- \\ \hline 
& Text & 0.8618 & 0.7327 & 0.6213 & 0.8052 & 0.7765 & 0.8752 & 0.7412 & 0.6254 & 0.8147 & 0.7825 & 0.7832 & 0.6622 & 0.5605 & 0.7276 & 0.7004 \\ 
& +Venue & {\cellcolor{blue!15} 0.8634} & {\cellcolor{blue!15} 0.7347} & {\cellcolor{blue!15} 0.6234} & {\cellcolor{blue!15} 0.8071} & {\cellcolor{blue!15} 0.7787} & {\cellcolor{blue!15} 0.8779} & {\cellcolor{blue!15} 0.7455} & {\cellcolor{blue!15} 0.6299} & {\cellcolor{blue!15} 0.8190} & {\cellcolor{blue!15} 0.7875} & {\cellcolor{blue!15} 0.7851} & {\cellcolor{blue!15} 0.6637} & {\cellcolor{blue!15} 0.5631} & {\cellcolor{blue!15} 0.7295} & {\cellcolor{blue!15} 0.7031} \\ 
& +Author & {\cellcolor{red!15} 0.8601} & {\cellcolor{red!15} 0.7319} & 0.6215 & {\cellcolor{red!15} 0.8036} & {\cellcolor{red!15} 0.7756} & {\cellcolor{red!15} 0.8641} & {\cellcolor{red!15} 0.7310} & {\cellcolor{red!15} 0.6171} & {\cellcolor{red!15} 0.8018} & {\cellcolor{red!15} 0.7693} & {\cellcolor{red!15} 0.7814} & {\cellcolor{red!15} 0.6600} & 0.5606 & {\cellcolor{red!15} 0.7251} & {\cellcolor{red!15} 0.6990} \\ 
\multirow{-4}{*}{Geology} & +Reference & {\cellcolor{red!15} 0.8432} & {\cellcolor{red!15} 0.7197} & {\cellcolor{red!15} 0.6138} & {\cellcolor{red!15} 0.7886} & {\cellcolor{red!15} 0.7622} & {\cellcolor{red!15} 0.8485} & {\cellcolor{red!15} 0.7162} & {\cellcolor{red!15} 0.6031} & {\cellcolor{red!15} 0.7848} & {\cellcolor{red!15} 0.7503} & -- & -- & -- & -- & -- \\ \hline 
& Text & 0.7406 & 0.6191 & 0.5326 & 0.6584 & 0.6192 & 0.7715 & 0.6368 & 0.5403 & 0.6796 & 0.6335 & 0.6174 & 0.5315 & 0.4641 & 0.5612 & 0.5334 \\ 
& +Venue & 0.7421 & {\cellcolor{blue!15} 0.6210} & {\cellcolor{blue!15} 0.5341} & {\cellcolor{blue!15} 0.6603} & {\cellcolor{blue!15} 0.6211} & {\cellcolor{blue!15} 0.7756} & {\cellcolor{blue!15} 0.6421} & {\cellcolor{blue!15} 0.5455} & {\cellcolor{blue!15} 0.6848} & {\cellcolor{blue!15} 0.6390} & 0.6187 & {\cellcolor{blue!15} 0.5351} & {\cellcolor{blue!15} 0.4689} & {\cellcolor{blue!15} 0.5643} & {\cellcolor{blue!15} 0.5376} \\ 
& +Author & 0.7408 & {\cellcolor{blue!15} 0.6214} & {\cellcolor{blue!15} 0.5346} & {\cellcolor{blue!15} 0.6603} & {\cellcolor{blue!15} 0.6212} & {\cellcolor{red!15} 0.7642} & {\cellcolor{red!15} 0.6294} & {\cellcolor{red!15} 0.5330} & {\cellcolor{red!15} 0.6718} & {\cellcolor{red!15} 0.6252} & 0.6175 & {\cellcolor{blue!15} 0.5347} & {\cellcolor{blue!15} 0.4687} & {\cellcolor{blue!15} 0.5637} & {\cellcolor{blue!15} 0.5372} \\ 
\multirow{-4}{*}{Mathematics} & +Reference & 0.7412 & {\cellcolor{blue!15} 0.6237} & {\cellcolor{blue!15} 0.5383} & {\cellcolor{blue!15} 0.6623} & {\cellcolor{blue!15} 0.6242} & {\cellcolor{red!15} 0.7324} & {\cellcolor{red!15} 0.5974} & {\cellcolor{red!15} 0.4999} & {\cellcolor{red!15} 0.6390} & {\cellcolor{red!15} 0.5895} & -- & -- & -- & -- & -- \\ \hline 
& Text & 0.8193 & 0.6638 & 0.5436 & 0.7746 & 0.7720 & 0.8670 & 0.6994 & 0.5696 & 0.8203 & 0.8157 & 0.7490 & 0.5935 & 0.4808 & 0.6932 & 0.6845 \\ 
& +Venue & {\cellcolor{blue!15} 0.8233} & {\cellcolor{blue!15} 0.6667} & {\cellcolor{blue!15} 0.5459} & {\cellcolor{blue!15} 0.7786} & {\cellcolor{blue!15} 0.7761} & {\cellcolor{blue!15} 0.8692} & {\cellcolor{blue!15} 0.7009} & {\cellcolor{blue!15} 0.5707} & {\cellcolor{blue!15} 0.8228} & {\cellcolor{blue!15} 0.8180} & 0.7496 & 0.5943 & {\cellcolor{blue!15} 0.4823} & 0.6946 & {\cellcolor{blue!15} 0.6867} \\ 
& +Author & 0.8203 & {\cellcolor{blue!15} 0.6659} & {\cellcolor{blue!15} 0.5459} & {\cellcolor{blue!15} 0.7766} & {\cellcolor{blue!15} 0.7745} & {\cellcolor{red!15} 0.8558} & {\cellcolor{red!15} 0.6879} & {\cellcolor{red!15} 0.5597} & {\cellcolor{red!15} 0.8069} & {\cellcolor{red!15} 0.8016} & {\cellcolor{red!15} 0.7469} & 0.5928 & 0.4812 & 0.6921 & 0.6843 \\ 
\multirow{-4}{*}{\begin{tabular}[c]{@{}c@{}}Materials\\ Science\end{tabular}} & +Reference & 0.8206 & {\cellcolor{blue!15} 0.6654} & {\cellcolor{blue!15} 0.5462} & {\cellcolor{blue!15} 0.7762} & {\cellcolor{blue!15} 0.7747} & {\cellcolor{red!15} 0.8510} & {\cellcolor{red!15} 0.6840} & {\cellcolor{red!15} 0.5567} & {\cellcolor{red!15} 0.8021} & {\cellcolor{red!15} 0.7970} & -- & -- & -- & -- & -- \\ \hline 
& Text & 0.8367 & 0.7228 & 0.6219 & 0.7597 & 0.7050 & 0.8739 & 0.7552 & 0.6490 & 0.7941 & 0.7368 & 0.7461 & 0.6399 & 0.5585 & 0.6730 & 0.6300 \\ 
& +Venue & {\cellcolor{blue!15} 0.8452} & {\cellcolor{blue!15} 0.7269} & {\cellcolor{blue!15} 0.6248} & {\cellcolor{blue!15} 0.7650} & {\cellcolor{blue!15} 0.7095} & {\cellcolor{blue!15} 0.8787} & {\cellcolor{blue!15} 0.7593} & {\cellcolor{blue!15} 0.6520} & {\cellcolor{blue!15} 0.7985} & {\cellcolor{blue!15} 0.7407} & {\cellcolor{blue!15} 0.7491} & {\cellcolor{blue!15} 0.6419} & {\cellcolor{blue!15} 0.5609} & {\cellcolor{blue!15} 0.6754} & {\cellcolor{blue!15} 0.6328} \\ 
& +Author & 0.8372 & 0.7235 & {\cellcolor{blue!15} 0.6236} & 0.7602 & 0.7061 & {\cellcolor{red!15} 0.8665} & {\cellcolor{red!15} 0.7472} & {\cellcolor{red!15} 0.6406} & {\cellcolor{red!15} 0.7857} & {\cellcolor{red!15} 0.7276} & {\cellcolor{red!15} 0.7415} & {\cellcolor{red!15} 0.6372} & {\cellcolor{red!15} 0.5566} & {\cellcolor{red!15} 0.6698} & {\cellcolor{red!15} 0.6270} \\ 
\multirow{-4}{*}{Physics} & +Reference & {\cellcolor{red!15} 0.8340} & {\cellcolor{red!15} 0.7197} & {\cellcolor{red!15} 0.6213} & {\cellcolor{red!15} 0.7563} & {\cellcolor{red!15} 0.7029} & {\cellcolor{red!15} 0.8682} & {\cellcolor{red!15} 0.7486} & {\cellcolor{red!15} 0.6423} & {\cellcolor{red!15} 0.7873} & {\cellcolor{red!15} 0.7294} & -- & -- & -- & -- & -- \\ \hline 
& Text & 0.8446 & 0.7503 & 0.6609 & 0.7804 & 0.7292 & 0.8764 & 0.7743 & 0.6798 & 0.8069 & 0.7524 & 0.7448 & 0.6517 & 0.5772 & 0.6804 & 0.6371 \\ 
& +Venue & {\cellcolor{blue!15} 0.8474} & {\cellcolor{blue!15} 0.7523} & {\cellcolor{blue!15} 0.6626} & {\cellcolor{blue!15} 0.7829} & {\cellcolor{blue!15} 0.7317} & {\cellcolor{blue!15} 0.8791} & {\cellcolor{blue!15} 0.7766} & {\cellcolor{blue!15} 0.6813} & {\cellcolor{blue!15} 0.8094} & {\cellcolor{blue!15} 0.7544} & {\cellcolor{blue!15} 0.7484} & {\cellcolor{blue!15} 0.6571} & {\cellcolor{blue!15} 0.5840} & {\cellcolor{blue!15} 0.6856} & {\cellcolor{blue!15} 0.6436} \\ 
& +Author & {\cellcolor{red!15} 0.8433} & {\cellcolor{red!15} 0.7495} & 0.6610 & {\cellcolor{red!15} 0.7793} & 0.7286 & {\cellcolor{red!15} 0.8652} & {\cellcolor{red!15} 0.7574} & {\cellcolor{red!15} 0.6600} & {\cellcolor{red!15} 0.7909} & {\cellcolor{red!15} 0.7332} & {\cellcolor{red!15} 0.7427} & 0.6528 & {\cellcolor{blue!15} 0.5806} & 0.6808 & 0.6393 \\ 
\multirow{-4}{*}{Biology} & +Reference & 0.8456 & 0.7505 & {\cellcolor{blue!15} 0.6618} & 0.7807 & 0.7300 & {\cellcolor{red!15} 0.8722} & {\cellcolor{red!15} 0.7668} & {\cellcolor{red!15} 0.6697} & {\cellcolor{red!15} 0.8000} & {\cellcolor{red!15} 0.7432} & -- & -- & -- & -- & -- \\ \hline 
& Text & 0.8448 & 0.7208 & 0.6211 & 0.7720 & 0.7310 & 0.8866 & 0.7658 & 0.6587 & 0.8187 & 0.7749 & 0.7480 & 0.6311 & 0.5334 & 0.6768 & 0.6312 \\ 
& +Venue & {\cellcolor{blue!15} 0.8485} & {\cellcolor{blue!15} 0.7229} & {\cellcolor{blue!15} 0.6229} & {\cellcolor{blue!15} 0.7756} & {\cellcolor{blue!15} 0.7347} & {\cellcolor{blue!15} 0.8896} & {\cellcolor{blue!15} 0.7668} & 0.6590 & {\cellcolor{blue!15} 0.8214} & {\cellcolor{blue!15} 0.7774} & {\cellcolor{blue!15} 0.7503} & {\cellcolor{blue!15} 0.6361} & {\cellcolor{blue!15} 0.5413} & {\cellcolor{blue!15} 0.6819} & {\cellcolor{blue!15} 0.6390} \\ 
& +Author & 0.8444 & 0.7220 & {\cellcolor{blue!15} 0.6232} & 0.7727 & 0.7324 & {\cellcolor{red!15} 0.8811} & {\cellcolor{red!15} 0.7577} & {\cellcolor{red!15} 0.6499} & {\cellcolor{red!15} 0.8105} & {\cellcolor{red!15} 0.7654} & 0.7480 & {\cellcolor{blue!15} 0.6346} & {\cellcolor{blue!15} 0.5403} & {\cellcolor{blue!15} 0.6794} & {\cellcolor{blue!15} 0.6366} \\ 
\multirow{-4}{*}{Chemistry} & +Reference & {\cellcolor{red!15} 0.8425} & {\cellcolor{red!15} 0.7182} & 0.6204 & {\cellcolor{red!15} 0.7694} & {\cellcolor{red!15} 0.7295} & {\cellcolor{red!15} 0.8810} & {\cellcolor{red!15} 0.7568} & {\cellcolor{red!15} 0.6484} & {\cellcolor{red!15} 0.8102} & {\cellcolor{red!15} 0.7649} & -- & -- & -- & -- & -- \\ \hline 
& Text & 0.7894 & 0.6491 & 0.5325 & 0.7335 & 0.7227 & 0.8421 & 0.6935 & 0.5669 & 0.7849 & 0.7711 & 0.6935 & 0.5744 & 0.4789 & 0.6458 & 0.6413 \\ 
& +Venue & {\cellcolor{blue!15} 0.7936} & {\cellcolor{blue!15} 0.6528} & {\cellcolor{blue!15} 0.5360} & {\cellcolor{blue!15} 0.7379} & {\cellcolor{blue!15} 0.7275} & {\cellcolor{blue!15} 0.8436} & {\cellcolor{blue!15} 0.6951} & {\cellcolor{blue!15} 0.5685} & {\cellcolor{blue!15} 0.7867} & {\cellcolor{blue!15} 0.7729} & {\cellcolor{blue!15} 0.7012} & {\cellcolor{blue!15} 0.5818} & {\cellcolor{blue!15} 0.4884} & {\cellcolor{blue!15} 0.6534} & {\cellcolor{blue!15} 0.6516} \\ 
& +Author & {\cellcolor{red!15} 0.7809} & {\cellcolor{red!15} 0.6428} & {\cellcolor{red!15} 0.5297} & {\cellcolor{red!15} 0.7246} & {\cellcolor{red!15} 0.7148} & {\cellcolor{red!15} 0.8291} & {\cellcolor{red!15} 0.6800} & {\cellcolor{red!15} 0.5565} & {\cellcolor{red!15} 0.7688} & {\cellcolor{red!15} 0.7549} & {\cellcolor{red!15} 0.6873} & {\cellcolor{red!15} 0.5727} & {\cellcolor{blue!15} 0.4804} & {\cellcolor{red!15} 0.6424} & 0.6397 \\ 
\multirow{-4}{*}{Medicine} & +Reference & {\cellcolor{blue!15} 0.7913} & {\cellcolor{blue!15} 0.6520} & {\cellcolor{blue!15} 0.5362} & {\cellcolor{blue!15} 0.7364} & {\cellcolor{blue!15} 0.7264} & 0.8394 & 0.6902 & {\cellcolor{red!15} 0.5633} & {\cellcolor{red!15} 0.7817} & {\cellcolor{red!15} 0.7672} & -- & -- & -- & -- & -- \\ \hline 
\end{NiceTabular}
}
\label{tab:performance}
\end{table*}

\begin{table*}[!h]
\centering
\renewcommand\arraystretch{0.9}
\caption{Macro average of P@$k$ and NDCG@$k$ over the 20 datasets. \colorbox{blue!15}{Blue}, \colorbox{red!15}{Red}, and ``--'': the same meaning as in Table \ref{tab:performance}.}
\vspace{-0.5em}
\scalebox{0.72}{
\begin{NiceTabular}{c|c|ccccc|ccccc|ccccc}[colortbl-like]
\hline
& \multirow{2}{*}{\textbf{Input}} & \multicolumn{5}{c|}{\textbf{Parabel \cite{prabhu2018parabel}}} & \multicolumn{5}{c|}{\textbf{Transformer \cite{xun2020correlation}}} & \multicolumn{5}{c}{\textbf{OAG-BERT \cite{liu2022oag}}} \\ \cline{3-17}
& & \textbf{P@1} & \textbf{P@3} & \textbf{P@5} & \textbf{N@3} & \textbf{N@5} & \textbf{P@1} & \textbf{P@3} & \textbf{P@5} & \textbf{N@3} & \textbf{N@5} & \textbf{P@1} & \textbf{P@3} & \textbf{P@5} & \textbf{N@3} & \textbf{N@5} \\ \hline
& Text & 0.7513 & 0.5811 & 0.4678 & 0.7076 & 0.6977 & 0.7510 & 0.5673 & 0.4507 & 0.6896 & 0.6712 & 0.6983 & 0.5354 & 0.4275 & 0.6549 & 0.6429 \\ & +Venue & {\cellcolor{blue!15} 0.7554} & {\cellcolor{blue!15} 0.5858} & {\cellcolor{blue!15} 0.4717} & {\cellcolor{blue!15} 0.7130} & {\cellcolor{blue!15} 0.7033} & {\cellcolor{blue!15} 0.7599} & {\cellcolor{blue!15} 0.5753} & {\cellcolor{blue!15} 0.4572} & {\cellcolor{blue!15} 0.6995} & {\cellcolor{blue!15} 0.6812} & {\cellcolor{blue!15} 0.7004} & {\cellcolor{blue!15} 0.5391} & {\cellcolor{blue!15} 0.4318} & {\cellcolor{blue!15} 0.6588} & {\cellcolor{blue!15} 0.6480} \\ 
& +Author & 0.7512 & {\cellcolor{blue!15} 0.5817} & {\cellcolor{blue!15} 0.4687} & 0.7079 & {\cellcolor{blue!15} 0.6983} & {\cellcolor{red!15} 0.7442} & {\cellcolor{red!15} 0.5594} & {\cellcolor{red!15} 0.4433} & {\cellcolor{red!15} 0.6809} & {\cellcolor{red!15} 0.6617} & 0.6984 & {\cellcolor{blue!15} 0.5374} & {\cellcolor{blue!15} 0.4305} & {\cellcolor{blue!15} 0.6569} & {\cellcolor{blue!15} 0.6461} \\ 
\multirow{-4}{*}{Macro Average} & +Reference & {\cellcolor{red!15} 0.7487} & 0.5809 & {\cellcolor{blue!15} 0.4689} & {\cellcolor{red!15} 0.7065} & 0.6977 & {\cellcolor{red!15} 0.7277} & {\cellcolor{red!15} 0.5469} & {\cellcolor{red!15} 0.4328} & {\cellcolor{red!15} 0.6652} & {\cellcolor{red!15} 0.6459} & -- & -- & -- & -- & -- \\ \hline 
\end{NiceTabular}
}
\vspace{-0.5em}
\label{tab:avg_p}
\end{table*}

\section{Experiments}
\subsection{Setup}
\noindent \textbf{Datasets and Compared Methods.} We have introduced the 20 datasets in Section \ref{sec:data}. For each of the three classifiers, we test its performance when using text only, text+venue, text+author, and text+reference in order to check the effect of each metadata type separately. (Recall that OAG-BERT cannot take references as metadata features, so it does not have the text+reference variant.) For detailed hyperparameter settings, one can refer to Appendix \ref{sec:app_hyper}.

\vspace{1mm}

\noindent \textbf{Evaluation Metrics.} Following previous studies on multi-label text classification \cite{liu2017deep,xun2020correlation,zhang2021match}, we adopt P@$k$ and NDCG@$k$ as evaluation metrics, where $k=1,3,5$. Given a paper $p$, let ${\rm rank}(i)$ be the index of the $i$-th highest predicted label according to each classifier, then
\begin{equation}
\small
\begin{split}
    {\rm P@}k &= \frac{1}{k} \sum_{i = 1}^k y_{p, {\rm rank}(i)}. \\
    {\rm DCG@}k = \sum_{i=1}^k \frac{y_{p, {\rm rank}(i)}}{\log(i+1)}, \ \ \ &{\rm NDCG@}k = \frac{{\rm DCG@}k}{\sum_{i=1}^{\min(k, ||\bmy_p||_0)}\frac{1}{\log(i+1)}}.
\end{split}
\end{equation}

\subsection{Overall Analysis}
Table \ref{tab:performance} shows the P@$k$ and NDCG@$k$ scores of the three classifiers on the 20 datasets. We run each experiment 5 times with the average score reported. We conduct two-tailed t-tests to check the statistical significance of metadata's effect. To be specific, given a field and a classifier, if a score is significantly improved (with p-value $<0.05$) after using a certain type of metadata in comparison with using text only, we mark the score as \colorbox{blue!15}{blue} in Table \ref{tab:performance}; if a score significantly deteriorates (with p-value $<0.05$), we mark the score as \colorbox{red!15}{red}.

From Table \ref{tab:performance}, we can observe that: (1) The effect of metadata varies remarkably across different fields. For instance, author information is significantly helpful in Computer Science and Engineering when Parabel is utilized as the classifier but becomes harmful in Biology and Medicine when the same classifier is adopted. References are useful in Business and Economics when Parabel is employed but become disadvantageous in the same fields if Transformer is the tagger.
(2) Venues are consistently beneficial to scientific literature tagging in almost all cases. To be specific, all 20 fields significantly benefit from venues\footnote{We say one field \textit{significantly benefits from} one type of metadata (when using a certain classifier) if at least one of the five metrics is significantly improved (with p-value $<0.05$) after incorporating that type of metadata.} when Parabel is used, 18 fields when Transformer is used, and 18 fields when OAG-BERT is used. In the remaining 4 cases where venue information is not beneficial, neither is it harmful.
(3) Authors are helpful in a majority of (15 and 17, respectively) fields when Parabel and OAG-BERT are the taggers but rarely help when Transformer is adopted. References are useful in even fewer cases. The overall performance of the three types of metadata is reflected by the macro average of P@$k$ and NDCG@$k$ scores over the 20 datasets, which are shown in Table \ref{tab:avg_p}. The reason why authors and references do not work in the Transformer classifier is that their embeddings need to be trained from scratch without good initialization. In contrast, leveraging venues only incurs a small number of additional parameters. Therefore, if one aims to utilize author and reference information, some embedding pre-training techniques \cite{zhang2021match} may help.

\subsection{Effect in Different Fields}
\begin{figure}[t]
\centering
\includegraphics[width=\linewidth]{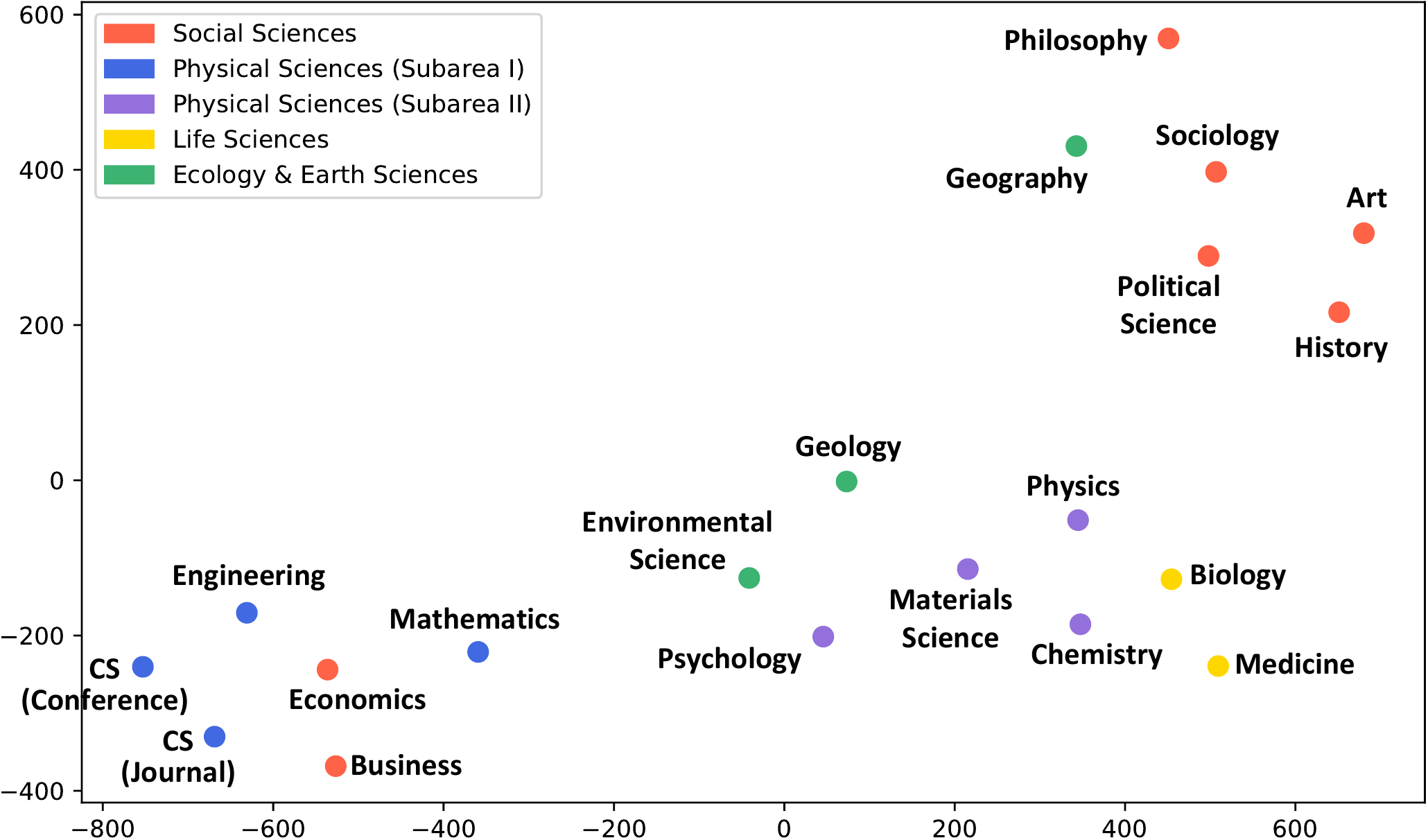}
\vspace{-1em}
\caption{We represent each field with a 24-dimensional vector based on the effect of venue, author, and reference information on the three classifiers. Then, we apply t-SNE \cite{maaten2008visualizing} to visualize these fields in a 2-dimensional space. The color scheme highlights several high-level scientific areas, following the major clusters of science detected by \cite{rosvall2011multilevel,yin2022public} and suggesting similar effects of metadata within each area.}
\vspace{-1em}
\label{fig:tsne}
\end{figure}

In this section, we examine the effect of metadata in different fields.
Given a field $F$, to describe the effect of metadata in $F$, we construct a 24-dimensional vector in the following way: There are three classifiers (i.e., Parabel, Transformer, and OAG-BERT) and three types of metadata (i.e., Venue, Author, and Reference) studied in our experiments, so there are $3\times3-1=8$ (classifier, metadata) combinations, since (OAG-BERT, Reference) is not applicable. For each of the 8 (classifier, metadata) combinations, we calculate the relative performance change of P@1, P@3, and P@5 by comparing the classifier using text only and the classifier using text together with the metadata. For example, given the (Parabel, Venue) combination, we calculate the following 3 values:
\begin{equation}
\small
\frac{{\rm P@}k({\rm Parabel, Text+Venue})-{\rm P@}k({\rm Parabel, Text})}{{\rm P@}k({\rm Parabel, Text})}, \ \ k=1,3,5.
\end{equation}
In total, we will have $3\times8=24$ values, which can form a 24-dimensional vector $\bmu_F$. We compute $\bmu_F$ for all 20 fields and then apply t-SNE \cite{maaten2008visualizing} to visualize these vectors in a 2-dimensional space. 

The visualization result is shown in Figure \ref{fig:tsne}, where we color each field according to high-level scientific areas/subareas detected by \cite{rosvall2011multilevel,yin2022public}. To be specific, Rosvall and Bergstrom \cite{rosvall2011multilevel} cluster scientific fields into four major areas -- social sciences, physical sciences, life sciences, and ecology \& earth sciences -- based on a large-scale paper citation network. The cluster of physical sciences is further split into two subareas, one of which is related to mathematics and computer science, and the other to physics and chemistry; Yin et al. \cite{yin2022public} observe a similar partitioning of the 19 fields when studying public uses of scientific papers in each field. 

In Figure \ref{fig:tsne}, we find that fields with the same color are often embedded closer, indicating that metadata have similar effects in fields belonging to the same area/subarea. This finding is very useful when we need to extrapolate the experience of using metadata in one field to a similar field. For example, one may have known that references are beneficial and authors are harmful when using Parabel in the Medicine field. Based on our finding, the same pattern can be deduced when using Parabel in the Biology field because Medicine and Biology are both life sciences. According to Table \ref{tab:performance}, this deduction is correct.

Meanwhile, we observe two exceptions: (1) The area of social sciences is split into two clusters, one of which contains Art, History, Philosophy, Sociology, and Political Science, and the other contains Economics and Business. The effects of metadata in the Business and Economics fields are more similar to those in Mathematics and Computer Science. This is possibly because a large proportion of Economics and Business papers rely on quantitative analysis of massive data to draw conclusions, which is different from the paradigm in most Art, History, and Philosophy papers. (2) Geography is embedded closer to social sciences than it is to Geology and Environmental Science. This is possibly because Geography is an interdisciplinary field. The physical geography subfield is more related to earth sciences while the human geography subfield is closer to social sciences. Overall, we still observe commonalities in the effects of metadata within high-level areas/subareas.

\subsection{Effect at Different Label Granularities}
\begin{figure}[t]
\raggedright
\subfloat[\rm Venue, Parabel]{
\includegraphics[width=0.15\textwidth]{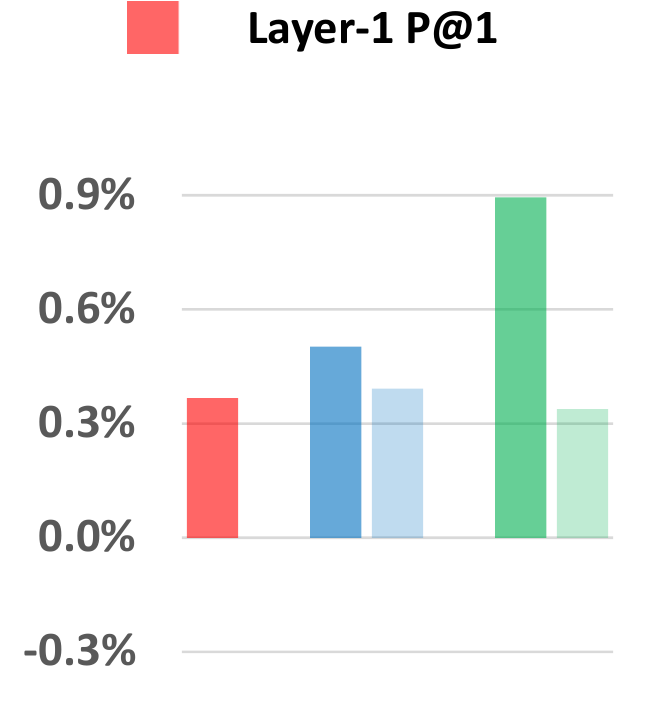}}
\hspace{-1mm}
\subfloat[\rm Venue, Transformer]{
\includegraphics[width=0.15\textwidth]{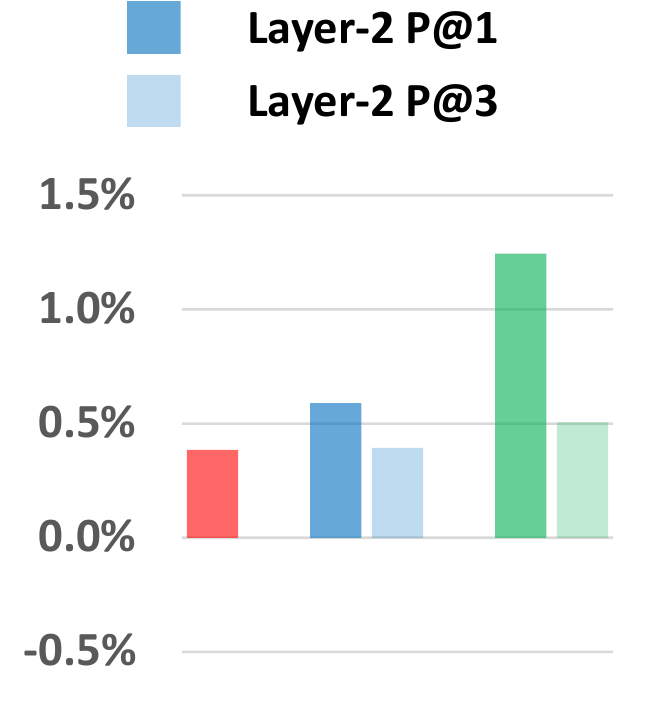}} 
\hspace{-1mm}
\subfloat[\rm Venue, OAG-BERT]{
\includegraphics[width=0.15\textwidth]{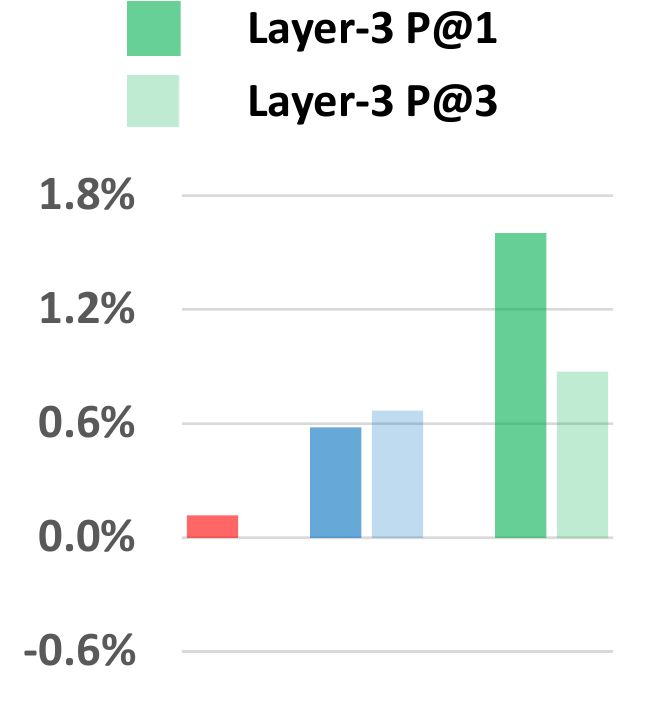}} \\
\subfloat[\rm Author, Parabel]{
\includegraphics[width=0.15\textwidth]{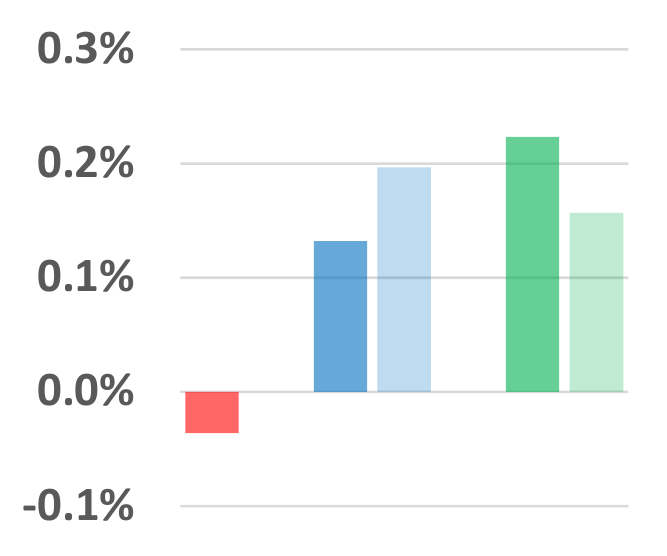}}
\hspace{-1mm}
\subfloat[\rm Author, Transformer]{
\includegraphics[width=0.15\textwidth]{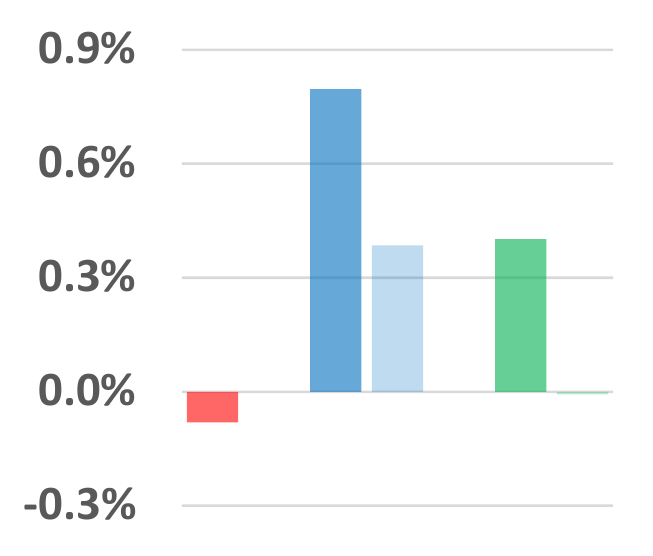}} 
\hspace{-1mm}
\subfloat[\rm Author, OAG-BERT]{
\includegraphics[width=0.15\textwidth]{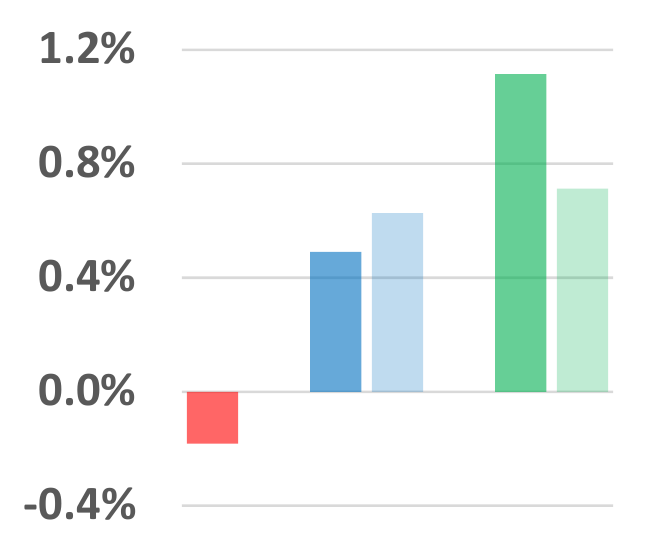}} \\
\subfloat[\rm Reference, Parabel]{
\includegraphics[width=0.15\textwidth]{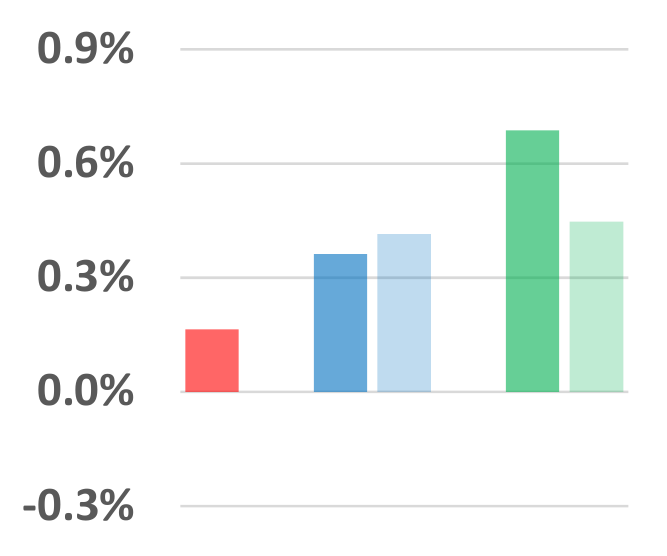}}
\hspace{-1mm}
\subfloat[\rm Reference, Transformer]{
\includegraphics[width=0.15\textwidth]{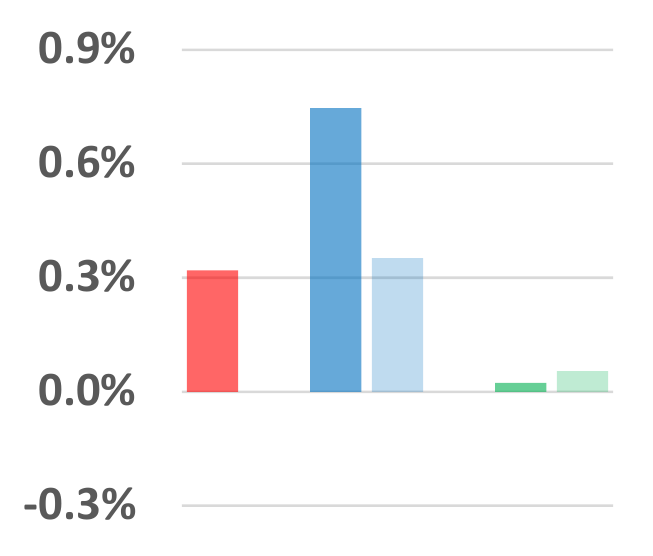}} \\
\vspace{-0.5em}
\caption{The effect of metadata on Layer-$j$ P@$k$ scores (averaged over the fields that significantly benefit from leveraging that type of metadata using the classifier).}
\vspace{-0.5em}
\label{fig:layer}
\end{figure}

\begin{table}[t]
\caption{The fields that benefit the most from leveraging venue information in terms of Layer-1 P@1, Layer-2 P@1, and Layer-3 P@1. ``T'': text only. ``+V'': text+venue. ``$\Delta$'': absolute performance improvement.}
\vspace{-0.5em}
\scalebox{0.68}{
\begin{tabular}{ccc|ccc|ccc}
\hline
     &    & \textbf{L1 P@1} &                                                                              &    & \textbf{L2 P@1} &                             &    & \textbf{L3 P@1} \\ \hline
\multicolumn{9}{c}{\textbf{Parabel \cite{prabhu2018parabel}}}                                                                                                                                                         \\ \hline
\multirow{3}{*}{Physics}     & T  & 0.7178 & \multirow{3}{*}{Philosophy}                                                  & T  & 0.7846 & \multirow{3}{*}{Philosophy} & T  & 0.6118 \\
     & +V & 0.7284 &                                                                              & +V & 0.8170 &                             & +V & 0.7209 \\
     & $\Delta$ & 1.06\% &                                                                              & $\Delta$ & 3.24\% &                             & $\Delta$ & 10.91\% \\ \hline
\multicolumn{9}{c}{\textbf{Transformer \cite{xun2020correlation}}}                                                                                                                                                     \\ \hline
\multirow{3}{*}{Physics}     & T  & 0.7393 & \multirow{3}{*}{Philosophy}                                                  & T  & 0.6728 & \multirow{3}{*}{Philosophy} & T  & 0.4105 \\
     & +V & 0.7495 &                                                                              & +V & 0.7504 &                             & +V & 0.6071 \\
     & $\Delta$ & 1.02\% &                                                                              & $\Delta$ & 7.76\% &                             & $\Delta$ & 19.66\% \\ \hline
\multicolumn{9}{c}{\textbf{OAG-BERT \cite{liu2022oag}}}                                                                                                                                                        \\ \hline
\multirow{3}{*}{Geography}   & T  & 0.6626 & \multirow{3}{*}{Art}                                                         & T  & 0.6747 & \multirow{3}{*}{History}    & T  & 0.3455 \\
     & +V & 0.6679 &                                                                              & +V & 0.6879 &                             & +V & 0.4130 \\
     & $\Delta$ & 0.53\% &                                                                              & $\Delta$ & 1.32\% &                             & $\Delta$ & 6.75\% \\ \hline
\end{tabular}
}
\vspace{-0.5em}
\label{tab:layer}
\end{table}

Now we examine the effect of metadata when predicting labels at different granularity levels. The MAG taxonomy \cite{shen2018web} has 6 layers (from the most coarse-grained Layer 0 to the most fine-grained Layer 5). Layer-0 labels are the 19 fields excluded from our label space. Across all 20 datasets, 85.2\%, 86.5\%, 70.4\%, 35.0\%, and 13.2\% of the papers have ground-truth Layer-1, Layer-2, Layer-3, Layer-4, and Layer-5 labels, respectively. Due to the low proportions of papers related to Layer-4 and Layer-5 tags, we only study the effect of metadata on predicting Layer-1, Layer-2, and Layer-3 tags.

Given one classifier and one type of metadata, we consider all fields that significantly benefit from the metadata using the classifier. In these fields, we calculate the Layer-$j$ P@$k$ scores ($(j,k)=(1,1), (2,1), (2,3), (3,1), (3,3)$). The definition of Layer-$j$ P@$k$ is very similar to that of P@$k$ except that Layer-$j$ P@$k$ considers the $k$ most confident labels at Layer $j$ instead of in the whole label space. When calculating Layer-$j$ P@$k$, we focus on papers with at least one ground-truth Layer-$j$ label. Then, we compute the absolute performance change of Layer-$j$ P@$k$ scores after incorporating the considered type of metadata. The results are shown in Figure \ref{fig:layer}.

From Figure \ref{fig:layer}, we find that: (1) On average, venues can help scientific literature tagging for not only coarse-grained tags but also fine-grained ones. This may be counterintuitive from computer scientists' perspective because CS venues (e.g., ``\textit{WWW}'') can hardly indicate very fine tags (e.g., ``\textsf{Link Farm}''). Indeed, in the Computer Science (Conference) dataset, the contribution of venues on Layer-3 P@1 is subtle (e.g., 0.19\% when using Parabel, -0.11\% when using Transformer). However, in fields other than Computer Science, some venues do carry very fine-grained signals. For example, there are two venues ``\textit{Journal of Roman Archaeology}'' and ``\textit{Mediaeval Studies}'' in History and Philosophy, respectively. These two venues may strongly imply a paper's relevance to ``\textsf{Classical Archaeology}'' and ``\textsf{Medievalism}'', which are Layer-2 and Layer-3 tags, respectively. 
In Table \ref{tab:layer}, we list the fields that benefit the most from leveraging venue information in terms of Layer-1 P@1, Layer-2 P@1, and Layer-3 P@1, where we do observe that Philosophy and History are the biggest beneficiaries in terms of Layer-3 P@1. (2) Different from venues, authors are beneficial to fine-grained tagging but harmful to coarse-grained prediction. This observation consistently holds across all three classifiers.

\section{Related Work}
\noindent \textbf{Extreme Multi-Label Text Classification.}
In keeping with our discussion in Section \ref{sec:model}, we divide related studies on extreme multi-label classification into three major categories. 
(1) \textit{Bag-of-words classifiers} \cite{prabhu2018parabel,babbar2017dismec,yen2017ppdsparse,prabhu2018extreme,jain2016extreme,prabhu2014fastxml,tagami2017annexml,yen2016pd,peng2016deepmesh,yu2022pecos,khandagale2020bonsai} take sparse tf--idf features as input. To improve model efficiency, 1-vs-all approaches such as DiSMEC \cite{babbar2017dismec} and PPDSparse \cite{yen2017ppdsparse} explore parallelism and model size reduction via model weight truncation. In another direction, tree-based approaches apply various partitioning techniques on the large label space. For example, Parabel \cite{prabhu2018parabel} partitions the labels to a balanced tree structure using 2-means clustering; Bonsai \cite{khandagale2020bonsai} improves Parabel by allowing multi-way and unbalanced partitions; XR-Linear \cite{yu2022pecos} improves Parabel by incorporating various hard negative sampling schemes; AnnexML \cite{tagami2017annexml} partitions the labels via graph-based nearest neighbor indices. 
(2) \textit{Sequence-based classifiers} \cite{liu2017deep,xun2019meshprobenet,you2019attentionxml,zhang2021match,ye2021beyond,xun2020correlation} employ deep neural architectures such as CNNs (e.g., XML-CNN \cite{liu2017deep}), RNNs (e.g., MeshProbeNet \cite{xun2019meshprobenet} and AttentionXML \cite{you2019attentionxml}), and Transformers (e.g., BertXML \cite{xun2020correlation}) to learn semantic representations of input text sequences for classification. There are also studies aggregating shallow word embeddings and/or applying MLP layers to obtain document embeddings, such as Slice \cite{jain2019slice}, DeepXML \cite{dahiya2021deepxml}, DECAF \cite{mittal2021decaf}, GalaXC \cite{saini2021galaxc}, and ECLARE \cite{mittal2021eclare}. 
(3) \textit{Pre-trained language model classifiers} \cite{chang2020taming,you2021bertmesh,zhang2022metadata,jiang2021lightxml,ye2020pretrained,zhang2021fast} propose to transfer the knowledge learned by PLMs from web-scale corpora to the classification task. For example, X-Transformer \cite{chang2020taming} complements the output of BERT \cite{devlin2019bert}, XLNet \cite{yang2019xlnet}, or RoBERTa \cite{liu2019roberta} with sparse tf--idf features; LightXML \cite{jiang2021lightxml} adopts PLMs as the text encoder and performs label shortlist and re-ranking with the same PLM; XR-Transformer \cite{zhang2021fast} further proposes fast multi-resolution PLM fine-tuning. Despite the success of these models in extreme multi-label classification, they mainly focus on classifying plain text sequences and are less aware of document metadata. In contrast, our work proposes several straightforward ways to enhance text classifiers with metadata signals.

\vspace{1mm}

\noindent \textbf{Scientific Literature Tagging.} Classifying academic papers is a common evaluation task in text mining \cite{cohan2020specter,zhang2022motifclass,mekala2020meta} and graph mining \cite{dong2017metapath2vec,hu2020heterogeneous} studies. However, most studies consider coarse-grained paper classification only (e.g., with 5-20 categories in the label space), the result of which is not subdivided enough to satisfy users' fine-grained interests. To tag papers on PubMed with fine-grained medical subject headings, the task of MeSH indexing has been extensively studied \cite{liu2015meshlabeler,peng2016deepmesh,xun2019meshprobenet,you2021bertmesh,dai2020fullmesh}. However, these models only use text or text+venue as input, leaving the effect of authors and references unexplored. Recently, Zhang et al. \cite{zhang2021match,zhang2022metadata} and Ye et al. \cite{ye2021beyond} make use of metadata to tag papers with fields of study in MAG. Still, these studies are restricted to computer science and biomedicine fields only. In comparison, our work conducts a systematic study across 19 fields and three major types of classifiers.

\section{Conclusions and Future Work}
In this work, we examine the effect of metadata on scientific literature tagging in 19 fields using three classifiers. 
Our results provide the following insights to practitioners aiming at building accurate scientific literature taggers: First, while previous metadata-aware approaches often directly use all types of available metadata, we show that not all of them are always beneficial. It is important to select useful metadata features based on the classifier's type, the field, and the granularity level of predicted tags. Second, although the state-of-the-art models for text sequence modeling have been dominated by Transformer-based models, we demonstrate that simple bag-of-words classifiers work comparably well or even better in many cases for large-scale fine-grained paper tagging, and may be more effective in leveraging different types of metadata. Third, despite the varying effects of different metadata types in different fields, the gain or loss induced by each metadata type is rather consistent across similar fields.


This study explores each type of metadata separately to avoid confounders. However, different types of metadata (e.g., a venue and an author) may interact with each other and provide additional hints for classification. In the future, it is of our interest to study the composite effect of multiple types of metadata.

\section*{Acknowledgments}
We thank anonymous reviewers for their valuable and insightful feedback.
Research was supported in part by the IBM-Illinois Discovery Accelerator Institute, US DARPA KAIROS Program No. FA8750-19-2-1004 and INCAS Program No. HR001121C0165, National Science Foundation IIS-19-56151, IIS-17-41317, and IIS 17-04532, and the Molecule Maker Lab Institute: An AI Research Institutes program supported by NSF under Award No. 2019897, and the Institute for Geospatial Understanding through an Integrative Discovery Environment (I-GUIDE) by NSF under Award No. 2118329. Any opinions, findings, and conclusions or recommendations expressed herein are those of the authors and do not necessarily represent the views, either expressed or implied, of DARPA or the U.S. Government.

\newpage
\appendix
\section{Appendix}
\begin{figure*}[t]
\centering
\includegraphics[width=\textwidth]{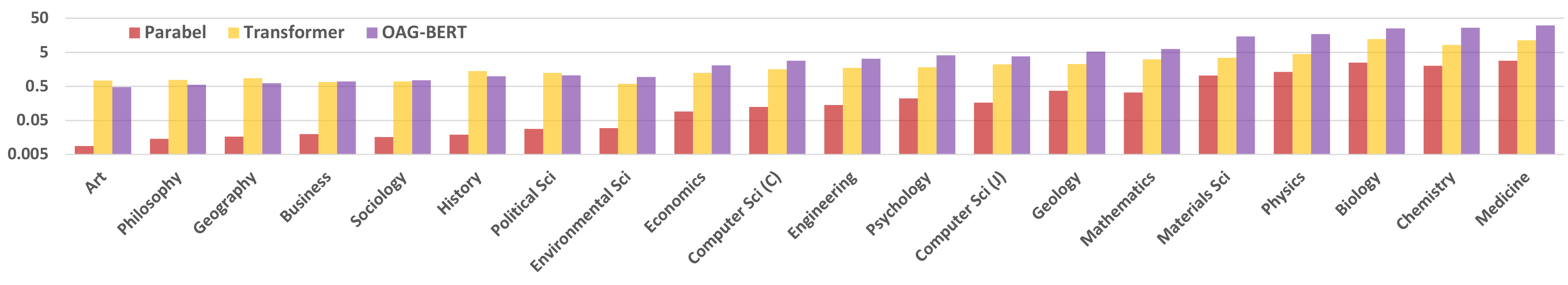}
\vspace{-2em}
\caption{Training time (in hours) of the three classifiers on the 20 datasets.}
\label{fig:time}
\end{figure*}

\subsection{Datasets}
\label{sec:app_data}
Supplementary to Table \ref{tab:dataset}, Table \ref{tab:app_dataset} summarizes more dataset statistics in \textsc{\dataset}, including the average numbers of authors, references, and labels per paper as well as the numbers of training, validation, and testing papers.

\begin{table}[!h]
\centering
\caption{More statistics of the 20 datasets in \textsc{\dataset}.}
\vspace{-0.5em}
\scalebox{0.68}{
\begin{NiceTabular}{c|ccc|ccc}
\hline
\textbf{Field} 
& \textbf{\begin{tabular}[c]{@{}c@{}}\#Authors/ \\ Paper\end{tabular}} 
& \textbf{\begin{tabular}[c]{@{}c@{}}\#References/ \\ Paper\end{tabular}} 
& \textbf{\begin{tabular}[c]{@{}c@{}}\#Labels/ \\ Paper\end{tabular}} 
& \textbf{\begin{tabular}[c]{@{}c@{}}\#Train \\ Papers\end{tabular}} 
& \textbf{\begin{tabular}[c]{@{}c@{}}\#Valid \\ Papers\end{tabular}} 
& \textbf{\begin{tabular}[c]{@{}c@{}}\#Test \\ Papers\end{tabular}} 
\\ \hline 
\rowcolor{black!15}
Art & 1.314         &  2.813           & 2.425             & 39,901                & 9,975             & 8,497            \\
Philosophy & 1.117         & 8.121           & 3.715             & 40,696                & 10,174             & 8,426            \\
\rowcolor{black!15}
Geography & 3.625         & 31.793         & 2.231             & 44,710                & 11,178            & 17,995           \\
Business & 2.346         & 37.635           & 3.556             & 49,481                & 12,370            & 23,007           \\
\rowcolor{black!15}
Sociology & 1.592         & 24.474          & 2.318          & 60,815                & 15,204             & 14,189            \\
History & 1.218         & 4.405         & 1.938            & 80,830                & 20,208            &  12,109           \\
\rowcolor{black!15}
\makecell{Political \\ Science} & 1.496         & 10.198          & 3.107           & 73,650                & 18,413             & 23,228             \\
\makecell{Environmental \\ Science} & 4.301         & 34.156         & 2.138               & 70,479               & 17,620            & 35,846           \\
\rowcolor{black!15}
Economics & 1.966         & 26.608         & 4.961             & 119,309                & 29,827            &  29,534          \\
Engineering & 3.055         & 19.703          & 4.987            & 179,804               & 44,951            & 45,251           \\
\rowcolor{black!15}
Psychology & 3.805         & 41.201         & 5.032            & 245,288               & 61,322            & 66,344          \\
CS (Conference) & 3.389      & 17.417          & 6.089           & 136,322                & 34,081           & 92,990           \\
CS (Journal) & 3.296         & 21.968          & 5.976           & 214,616                & 53,654           & 142,333          \\
\rowcolor{black!15}
Geology & 3.552         & 36.510           & 5.957            & 284,022               & 71,006            & 76,806          \\
Mathematics & 2.151         & 19.068          & 6.433           & 338,454                & 84,613            & 67,484          \\
\rowcolor{black!15}
\makecell{Materials \\ Science} & 4.829         & 26.838         & 4,428             & 783,289                & 195,822         & 358,620           \\
Physics & 9.841         & 23.509         & 7.271          & 905,613                & 226,403          & 237,967         \\
\rowcolor{black!15}
Biology & 5.611         & 36.137         & 8.306          & 1,125,605               & 281,401          & 181,772          \\
Chemistry & 4.293         & 28.382         & 6.288           & 1,252,531               & 313,133          & 284,292          \\
\rowcolor{black!15}
Medicine & 5.647         & 12.888        & 5.376           & 1,620,165               & 405,041          & 620,899          \\ \hline
\end{NiceTabular}
}
\label{tab:app_dataset}
\vspace{-0.5em}
\end{table}

\subsection{Hyperparameters}
\label{sec:app_hyper}
In each of the 20 datasets, we remove metadata instances that appear in less than 5 papers. We adopt the same hyperparameter configuration when using text only, text+venue, text+author, and text+reference as input. The code source and hyperparameters of each classifier are introduced below.

\subsubsection{Parabel}\footnote{\url{http://manikvarma.org/code/Parabel/download.html}} 
We remove words appearing in less than 5 papers. All parameters are set by default. Specifically, the number of threads $T=1$; the number of trees $t=3$; the maximum number of labels in a leaf node $m=100$; the beam search width in prediction $B=10$.

\subsubsection{Transformer}\footnote{\url{https://github.com/XunGuangxu/CorNet} \ \ \ \ \ (We use the BertXML classifier in this GitHub repository. Although the model is called BertXML, it trains a Transformer architecture from scratch without BERT initialization.)} 
The number of Transformer layers is 1; the number of attention heads is 2; the number of [CLS] tokens $C=10$; the embedding dimension is 100; the maximum sequence length is 500; the batch size is 256. We adopt GloVe.6B.100d as initialized word embeddings. For (the number of training epochs, the number of warm-up epochs), we use (100, 20) for Art, Philosophy, and Geography, (80, 16) for Business, Sociology, History, and Political Science, (60, 12) for Environmental Science, Economics, Engineering, and Computer Science, (40, 8) for Psychology, Geology, and Mathematics, (20, 4) for Materials Science, Physics, and Biology, and (15, 3) for Chemistry and Medicine. Other hyperparameters are set by default.

\subsubsection{OAG-BERT}\footnote{\url{https://github.com/THUDM/OAG-BERT}}
We use ``oagbert-v2'' as the PLM. After PLM encoding, we fix paper embeddings to train a Parabel classifier. All parameters of Parabel are set by default.

\subsection{More on the Effect of Metadata at Different Label Granularities}
Supplementary to Table \ref{tab:layer}, Table \ref{tab:app_layer} shows the fields that benefit the most from leveraging author or reference information in terms of Layer-1 P@1, Layer-2 P@1, and Layer-3 P@1. We find that authors and references are beneficial to the Art, Philosophy, and History fields in many cases. The possible reason is that each paper in these fields has a small number of authors and references (according to the statistics in Table \ref{tab:app_dataset}). Therefore, the author or reference list may contain fewer confounding signals and be more topic-indicative. 

\begin{table}[!h]
\caption{The fields that benefit the most from leveraging author or reference information in terms of Layer-1 P@1, Layer-2 P@1, and Layer-3 P@1. ``T'': text only. ``+A'': text+author. ``+R'': text+reference. ``$\Delta$'': absolute performance improvement.}
\vspace{-0.5em}
\scalebox{0.68}{
\begin{tabular}{ccc|ccc|ccc}
\hline
     &    & \textbf{L1 P@1} &                                                                              &    & \textbf{L2 P@1} &                             &    & \textbf{L3 P@1} \\ \hline
\multicolumn{9}{c}{\textbf{Parabel \cite{prabhu2018parabel}}}                                                                                                                                                         \\ \hline
\multirow{3}{*}{Mathematics} & T  & 0.5960 & \multirow{3}{*}{\begin{tabular}[c]{@{}c@{}}Political\\ Science\end{tabular}} & T  & 0.7928 & \multirow{3}{*}{Philosophy} & T  & 0.6118 \\
     & +A & 0.5971 &                                                                              & +A & 0.7969 &                             & +A & 0.6160 \\
     & $\Delta$ & 0.11\% &                                                                              & $\Delta$ & 0.41\% &                             & $\Delta$ & 0.42\% \\ \hline
\multirow{3}{*}{Psychology}  & T  & 0.6992 & \multirow{3}{*}{Business}                                                    & T  & 0.6704 & \multirow{3}{*}{Business}   & T  & 0.6124 \\
     & +R & 0.7086 &                                                                              & +R & 0.6801 &                             & +R & 0.6265 \\
     & $\Delta$ & 0.94\% &                                                                              & $\Delta$ & 0.97\% &                             & $\Delta$ & 1.41\% \\ \hline
\multicolumn{9}{c}{\textbf{Transformer \cite{xun2020correlation}}}                                                                                                                                                     \\ \hline
\multirow{3}{*}{History}     & T  & 0.5471 & \multirow{3}{*}{Philosophy}                                                  & T  & 0.6728 & \multirow{3}{*}{Art}        & T  & 0.4859 \\
     & +A & 0.5489 &                                                                              & +A & 0.6900 &                             & +A & 0.4910 \\
     & $\Delta$ & 0.18\% &                                                                              & $\Delta$ & 1.72\% &                             & $\Delta$ & 0.51\% \\ \hline
\multirow{3}{*}{History}     & T  & 0.5471 & \multirow{3}{*}{Art}                                                         & T  & 0.7859 & \multirow{3}{*}{History}    & T  & 0.4857 \\
     & +R & 0.5537 &                                                                              & +R & 0.7953 &                             & +R & 0.4890 \\
     & $\Delta$ & 0.66\% &                                                                              & $\Delta$ & 0.94\% &                             & $\Delta$ & 0.33\% \\ \hline
\multicolumn{9}{c}{\textbf{OAG-BERT \cite{liu2022oag}}}                                                                                                                                                        \\ \hline
\multirow{3}{*}{Art}         & T  & 0.6241 & \multirow{3}{*}{Art}                                                         & T  & 0.6747 & \multirow{3}{*}{History}    & T  & 0.3455 \\
     & +A & 0.6254 &                                                                              & +A & 0.6875 &                             & +A & 0.3819 \\
     & $\Delta$ & 0.13\% &                                                                              & $\Delta$ & 1.28\% &                             & $\Delta$ & 3.64\% \\ \hline
\end{tabular}
}
\vspace{-0.5em}
\label{tab:app_layer}
\end{table}

\subsection{Effect of Metadata on Efficiency}
Now we report the effect of metadata on model efficiency. Table \ref{tab:time} shows the average relative training time increase of the three classifiers across the 20 datasets after incorporating venues, authors, and references, respectively. All models are run on Intel Xeon E5-2680 v2 @ 2.80GHz and one NVIDIA GeForce GTX 1080 Ti GPU (if a GPU is needed). We can observe a significant increase in training time after adding references as features. When Parabel is the classifier, the increase is due to a much longer bag-of-words vector used to represent a paper; when Transformer is the classifier, the increase is caused by a large number of additional parameters (i.e., reference embeddings), which make the model converge more slowly. 

\begin{table}[!h]
\centering
\caption{Average relative increase in training time across 20 datasets after incorporating one type of metadata.}
\vspace{-0.5em}
\scalebox{0.72}{
\begin{tabular}{c|ccc}
\hline
\textbf{}  & \textbf{Parabel \cite{prabhu2018parabel}} & \textbf{Transformer \cite{xun2020correlation}} & \textbf{OAG-BERT \cite{liu2022oag}} \\ \hline
+Venue     & +0.15\%          &     +0.13\%          & +0.11\%           \\
+Author    & +0.72\%          &     +1.86\%          & +0.11\%           \\
+Reference & +22.68\%         &     +10.75\%         & --                 \\ \hline
\end{tabular}
}
\vspace{-0.5em}
\label{tab:time}
\end{table}

\begin{table*}[t]
\centering
\caption{Statistics of the three additional datasets with MeSH labels.}
\vspace{-0.5em}
\scalebox{0.68}{
\begin{NiceTabular}{c|cccccc|ccc|ccc}
\hline
\textbf{Field} 
& \textbf{\begin{tabular}[c]{@{}c@{}}Paper\\ Source\end{tabular}} 
& \textbf{\#Papers} 
& \textbf{\#Labels}  
& \textbf{\#Venues} 
& \textbf{\#Authors}
& \textbf{\#References}
& \textbf{\begin{tabular}[c]{@{}c@{}}\#Authors/ \\ Paper\end{tabular}} 
& \textbf{\begin{tabular}[c]{@{}c@{}}\#References/ \\ Paper\end{tabular}} 
& \textbf{\begin{tabular}[c]{@{}c@{}}\#Labels/ \\ Paper\end{tabular}} 
& \textbf{\begin{tabular}[c]{@{}c@{}}\#Train \\ Papers\end{tabular}} 
& \textbf{\begin{tabular}[c]{@{}c@{}}\#Valid \\ Papers\end{tabular}} 
& \textbf{\begin{tabular}[c]{@{}c@{}}\#Test \\ Papers\end{tabular}}
\\ \hline 
\rowcolor{black!15}
Biology-MeSH & Journal & 1,379,393 & 25,039 & 100 & 2,486,814 & 6,876,739 & 5.686 & 40.043 & 13.870 & 985,364  & 246,341 & 147,688 \\
Chemistry-MeSH & Journal & 762,129  & 21,585 & 87  & 1,498,358 & 5,928,908 & 4.741 & 34.344 & 10.984 & 511,814  & 127,954 & 122,361 \\
\rowcolor{black!15}
Medicine-MeSH & Journal & 1,536,660 & 25,188 & 100 & 2,791,165 & 7,190,021 & 5.254 & 20.931 & 11.819 & 1,020,969 & 255,242 & 260,449 \\ \hline
\end{NiceTabular}
}
\label{tab:dataset_mesh}
\end{table*}

\begin{table*}[!h]
\centering
\renewcommand\arraystretch{0.9}
\caption{P@$k$ and NDCG@$k$ scores on the three additional datasets. \colorbox{blue!15}{Blue}, \colorbox{red!15}{Red}, and ``--'': the same meaning as in Table \ref{tab:performance}.}
\vspace{-0.5em}
\scalebox{0.72}{
\begin{NiceTabular}{c|c|ccccc|ccccc|ccccc}[colortbl-like]
\hline
\multirow{2}{*}{\textbf{Field}} & \multirow{2}{*}{\textbf{Input}} & \multicolumn{5}{c|}{\textbf{Parabel \cite{prabhu2018parabel}}} & \multicolumn{5}{c|}{\textbf{Transformer \cite{xun2020correlation}}} & \multicolumn{5}{c}{\textbf{OAG-BERT \cite{liu2022oag}}} \\ \cline{3-17}
 & & \textbf{P@1} & \textbf{P@3} & \textbf{P@5} & \textbf{N@3} & \textbf{N@5} & \textbf{P@1} & \textbf{P@3} & \textbf{P@5} & \textbf{N@3} & \textbf{N@5} & \textbf{P@1} & \textbf{P@3} & \textbf{P@5} & \textbf{N@3} & \textbf{N@5} \\ \hline
& Text & 0.8924 & 0.7964 & 0.7088 & 0.8194 & 0.7576 & 0.9112 & 0.8098 & 0.7193 & 0.8341 & 0.7698 & 0.7506 & 0.6185 & 0.5460 & 0.6490 & 0.5945 \\ 
& +Venue & 0.8934 & {\cellcolor{blue!15} 0.7976} & {\cellcolor{blue!15} 0.7101} & {\cellcolor{blue!15} 0.8206} & {\cellcolor{blue!15} 0.7587} & 0.9119 & 0.8105 & 0.7194 & 0.8348 & 0.7701 & {\cellcolor{blue!15} 0.7520} & 0.6200 & 0.5473 & 0.6504 & 0.5958 \\ 
& +Author & 0.8932 & {\cellcolor{blue!15} 0.7985} & {\cellcolor{blue!15} 0.7112} & {\cellcolor{blue!15} 0.8212} & {\cellcolor{blue!15} 0.7596} & {\cellcolor{red!15} 0.9090} & {\cellcolor{red!15} 0.8028} & {\cellcolor{red!15} 0.7093} & {\cellcolor{red!15} 0.8281} & {\cellcolor{red!15} 0.7614} & 0.7504 & 0.6184 & 0.5464 & 0.6488 & 0.5946 \\ 
\multirow{-4}{*}{Biology-MeSH} & +Reference & {\cellcolor{blue!15} 0.8976} & {\cellcolor{blue!15} 0.8058} & {\cellcolor{blue!15} 0.7198} & {\cellcolor{blue!15} 0.8279} & {\cellcolor{blue!15} 0.7674} & {\cellcolor{red!15} 0.9079} & {\cellcolor{red!15} 0.7988} & {\cellcolor{red!15} 0.7034} & {\cellcolor{red!15} 0.8248} & {\cellcolor{red!15} 0.7565} & --      & --      & --      & --      & --       \\ \hline 
& Text & 0.8447 & 0.7340 & 0.6407 & 0.7603 & 0.6947 & 0.8445 & 0.7113 & 0.6082 & 0.7427 & 0.6684 & 0.6971 & 0.5804 & 0.5099 & 0.6071 & 0.5557 \\ 
& +Venue & 0.8453 & {\cellcolor{blue!15} 0.7354} & {\cellcolor{blue!15} 0.6421} & {\cellcolor{blue!15} 0.7615} & {\cellcolor{blue!15} 0.6960} & 0.8465 & {\cellcolor{blue!15} 0.7151} & {\cellcolor{blue!15} 0.6121} & {\cellcolor{blue!15} 0.7460} & {\cellcolor{blue!15} 0.6720} & {\cellcolor{blue!15} 0.6995} & 0.5826 & {\cellcolor{blue!15} 0.5132} & 0.6093 & {\cellcolor{blue!15} 0.5587} \\ 
& +Author & 0.8450 & {\cellcolor{blue!15} 0.7350} & {\cellcolor{blue!15} 0.6419} & {\cellcolor{blue!15} 0.7611} & {\cellcolor{blue!15} 0.6957} & 0.8439 & 0.7105 & 0.6066 & 0.7419 & 0.6670 & 0.6981 & 0.5819 & 0.5126 & 0.6086 & 0.5580 \\ 
\multirow{-4}{*}{Chemistry-MeSH} & +Reference & {\cellcolor{blue!15} 0.8490} & {\cellcolor{blue!15} 0.7409} & {\cellcolor{blue!15} 0.6491} & {\cellcolor{blue!15} 0.7667} & {\cellcolor{blue!15} 0.7023} & {\cellcolor{red!15} 0.8248} & {\cellcolor{red!15} 0.6802} & {\cellcolor{red!15} 0.5743} & {\cellcolor{red!15} 0.7139} & {\cellcolor{red!15} 0.6366} & --      & --      & --      & --      & --      \\ \hline 
& Text & 0.9673 & 0.8410 & 0.7454 & 0.8712 & 0.8075 & 0.9683 & 0.8567 & 0.7583 & 0.8842 & 0.8191 & 0.7343 & 0.6229 & 0.5629 & 0.6491 & 0.6089 \\ 
& +Venue & 0.9673 & {\cellcolor{blue!15} 0.8422} & {\cellcolor{blue!15} 0.7472} & {\cellcolor{blue!15} 0.8722} & {\cellcolor{blue!15} 0.8090} & 0.9688 & {\cellcolor{blue!15} 0.8585} & {\cellcolor{blue!15} 0.7606} & {\cellcolor{blue!15} 0.8856} & {\cellcolor{blue!15} 0.8210} & {\cellcolor{blue!15} 0.7370} & 0.6235 & 0.5607 & 0.6503 & 0.6081 \\ 
& +Author & 0.9671 & {\cellcolor{red!15} 0.8342} & {\cellcolor{red!15} 0.7424} & {\cellcolor{red!15} 0.8656} & {\cellcolor{red!15} 0.8039} & 0.9687 & {\cellcolor{red!15} 0.8495} & {\cellcolor{red!15} 0.7496} & {\cellcolor{red!15} 0.8786} & {\cellcolor{red!15} 0.8117} & 0.7347 & 0.6185 & {\cellcolor{red!15} 0.5549} & 0.6459 & {\cellcolor{red!15} 0.6027} \\ 
\multirow{-4}{*}{Medicine-MeSH} & +Reference & {\cellcolor{red!15} 0.9666} & {\cellcolor{red!15} 0.8382} & {\cellcolor{blue!15} 0.7463} & {\cellcolor{red!15} 0.8687} & 0.8073 & {\cellcolor{red!15} 0.9662} & {\cellcolor{red!15} 0.8470} & {\cellcolor{red!15} 0.7458} & {\cellcolor{red!15} 0.8762} & {\cellcolor{red!15} 0.8084} & --      & --      & --      & --      & -- \\ \hline 
\end{NiceTabular}
}
\label{tab:performance_mesh}
\end{table*}

Figure \ref{fig:time} shows the training time of the three classifiers on the 20 datasets. The reported training time is an average over 20 runs (i.e., 5 runs $\times$ $\{$text only, text+venue, text+author, text+reference$\}$) when Parabel and Transformer are tested, or 15 runs (i.e., 5 runs $\times$ $\{$text only, text+venue, text+author$\}$) when OAG-BERT is tested. Among the three classifiers, Parabel is consistently the most efficient across the 20 datasets; Transformer spends more time than OAG-BERT on small datasets, but the situation is reversed on moderate-sized and large datasets.

\subsection{Additional Datasets with MeSH Labels}
\label{sec:app_mesh}
To further strengthen our findings, we construct three datasets with Medical Subject Headings (MeSH) terms \cite{coletti2001medical} as their labels, which are curated by experts from the National Library of Medicine. To be specific, we take the three datasets, Biology, Chemistry, and Medicine, from \textsc{\dataset} and obtain the ground-truth MeSH labels of each paper\footnote{\url{https://learn.microsoft.com/en-us/academic-services/graph/reference-data-schema\#paper-mesh}}. After removing papers not having MeSH labels, we get three new datasets, Biology-MeSH, Chemistry-MeSH, and Medicine-MeSH. Their statistics are shown in Table \ref{tab:dataset_mesh}. 

We run Parabel, Transformer, and OAG-BERT on the three new datasets. When running Transformer, for (the number of training epochs, the number of warm-up epochs), we use (20, 4) for all three datasets. When running OAG-BERT, we need to rerank those labels appearing in the paper text higher than those not. Note that a MeSH label may have multiple label names (i.e., one canonical name and 0, 1, or several entry terms, see the MeSH label ``\textsf{COVID-19}''\footnote{\url{https://meshb.nlm.nih.gov/record/ui?ui=D000086382}} as an example). Given a MeSH label, if any of its label names appears in the paper text, we view it as occurring in the paper. All other hyperparameters are the same as in Appendix \ref{sec:app_hyper}.

The P@$k$ and NDCG@$k$ scores of Parabel, Transformer, and OAG-BERT on the three new datasets are demonstrated in Table \ref{tab:performance_mesh}. In general, we still find that venues are beneficial to scientific literature tagging in almost all cases, while the effect of authors and references depends on the classifier's type and the field.

The three additional datasets are also available in our \textsc{\dataset} benchmark: {\color{myblue} \url{https://doi.org/10.5281/zenodo.7611544}}.
\end{spacing}

\balance
\bibliographystyle{ACM-Reference-Format}
\bibliography{www}
\end{document}